%% file: EMDLOT3.tex
\documentclass[12pt]{article}
\usepackage{amsmath,amsfonts,graphicx,rotating,afterpage}
\usepackage{amssymb,  epsfig}
\usepackage{psfrag}
\usepackage{epstopdf}
\usepackage[usenames]{color}
\usepackage{float}
\usepackage{makecell}
\usepackage{bm}
\usepackage{CJKutf8}
\usepackage{amsfonts}
\usepackage[title]{appendix}
\usepackage{authblk}
\usepackage{tabularx}
\usepackage{makecell}
\usepackage{geometry}
\usepackage{changepage}
\usepackage{booktabs}
\usepackage{array}
\usepackage{url}
\usepackage{csquotes}
\usepackage{booktabs} 
\usepackage{textcomp}
\usepackage[round, sort]{natbib}
\usepackage{setspace}
\usepackage[colorlinks,linkcolor=blue,       
            anchorcolor=blue,  
            citecolor=blue]{hyperref}

\def\be{\begin{equation}}
\def\ee{\end{equation}}

\begin{document}

\title{\large \textbf{Why Bonds Fail Differently? Explainable Multimodal Learning for Multi-Class Default Prediction}
}
\author{}
\author{\small Yi Lu\thanks{School of Economics and Finance, Shanghai International Studies University, Shanghai, 200083, China, E-mail: 0243700766@shisu.edu.cn. },
 \quad\small Aifan Ling\thanks{Corresponding author. School of Economics and Finance, Shanghai International Studies University, Shanghai, 201620, China, E-mail: aiffling@163.com. },
 \quad\small Chaoqun Wang\thanks{School of AI and Advanced Computing, Xi'an Jiaotong-Liverpool University, Suzhou, 215400, China, E-mail: Chaoqun.Wang@xjtlu.edu.cn. },
 \quad\small Yaxin Xu\thanks{School of Foreign Studies, Shanghai University of Finance and Economics, Shanghai, 200433, China, E-mail: 2024214242@stu.sufe.edu.cn },
 }

\maketitle

\begin{abstract}
In recent years, China's bond market has seen a surge in defaults amid regulatory reforms and macroeconomic volatility. Traditional machine learning models struggle to capture financial data's irregularity and temporal dependencies, while most deep learning models lack interpretability-critical for financial decision-making. To tackle these issues, we propose EMDLOT (Explainable Multimodal Deep Learning for Time-series), a novel framework for multi-class bond default prediction. EMDLOT integrates numerical time-series (financial/macroeconomic indicators) and unstructured textual data (bond prospectuses), uses Time-Aware LSTM to handle irregular sequences, and adopts soft clustering and multi-level attention to boost interpretability. Experiments on 1994 Chinese firms (2015-2024) show EMDLOT outperforms traditional (e.g., XGBoost) and deep learning (e.g., LSTM) benchmarks in recall, F1-score, and mAP, especially in identifying default/extended firms. Ablation studies validate each component's value, and attention analyses reveal economically intuitive default drivers. This work provides a practical tool and a trustworthy framework for transparent financial risk modeling.


\vskip5pt

\bigskip

\vskip5pt%


\thispagestyle{empty} 


\end{abstract}


\baselineskip21pt 

\onehalfspacing

\section{Introduction}
\label{sec1}

Since the phasing out of China's bond bailout policy in 2014, the country's bond market has undergone a paradigm shift, transitioning into a mature credit market characterized by a marked uptick in both the frequency and scale of bond defaults \citep{mo2021china}. In recent years, China's bond market has risen to become the world's second-largest, with its total size surpassing 100 trillion RMB. Nevertheless, it has been plagued by a resurgence of default incidents, especially amid the economic headwinds spanning 2020 to 2023. A case in point is the debt crisis of Evergrande Group during 2021-2022, which garnered global attention. Notably, the spillover effects of its credit defaults were more pronounced in the credit market than in the equity market \citep{altman2022has}.

In the field of bond default prediction, traditional machine learning models have long focused on static feature analysis, rendering them inadequate in capturing long-term temporal dependencies inherent in time-series data. In contrast, deep learning has emerged as a potent tool for streamlining complex data processing tasks, encompassing classification, regression, and generative modeling \citep{dell2025deep}. However, a critical challenge persists: due to data gaps and non-mandatory disclosure requirements imposed by exchanges, the financial indicators of numerous enterprises exhibit varying lengths and irregular time intervals. Most deep learning models, which demand fixed intervals and step sizes, struggle to accommodate such irregularities, severely undermining their adaptability and practical utility in real-world applications.
Within the financial sector, the majority of deep learning models prioritize predictive accuracy at the expense of interpretability. Yet, interpretability has increasingly become a cornerstone of financial systems, as it is pivotal for ensuring transparency, fostering trust, and facilitating informed decision-making in intricate financial landscapes. Currently, the financial industry relies heavily on post hoc interpretability techniques, such as LIME and SHAP \citep{mohsin2025explaining}, while the models themselves often lack intrinsic interpretability-a gap that hinders their widespread adoption and regulatory acceptance.

In multimodal prediction, textual information plays a pivotal role, and natural language processing (NLP) has proven instrumental in enhancing the efficiency and accuracy of business analyses \citep{ong2025explainable}. However, conventional frequency-based textual analysis models are plagued by high-dimensional sparsity and fail to capture the subtle nuances of sentiment in financial texts \citep{tan2023survey}. Furthermore, existing Chinese sentiment dictionaries exhibit significant biases when applied to contexts related to financial stability, limiting their effectiveness in bond default prediction scenarios.

Additionally, a fundamental flaw in current research lies in the framing of bond default prediction as a binary classification task-simply distinguishing between "normal" and "default" states. This oversimplification fails to account for the granularity of different default types, misaligning with the complexities of real-world financial scenarios where varying default severities and mechanisms demand more nuanced categorization.

Against this backdrop, three innovative research questions emerge: First, can a novel deep learning architecture be developed to dynamically adapt to irregular time intervals in financial time-series data, thereby improving the robustness of bond default prediction? Second, how can intrinsic interpretability be integrated into multimodal models that combine financial time-series and textual sentiment data, addressing the "black box" problem while maintaining predictive performance? Third, can a multi-class classification framework be established to categorize bond defaults based on severity and underlying causes, bridging the gap between academic models and practical risk management needs?

Inspired by these limitations and problems, this paper proposes EMDLOT (Explainable Multimodal Deep Learning for Time-series), a model capable of handling three-class bond default prediction tasks. In terms of multimodality, the model integrates numerical and textual modalities, selecting corporate financial indicators, macroeconomic indicators, and bond prospectuses as multimodal data sources to comprehensively capture the interplay between corporate operational conditions and external risk signals. For temporal modeling, the model employs Time-Aware LSTM, which not only accounts for the dynamic evolution of corporate financial indicators over time but also incorporates time intervals as inputs. This design enables the model to adapt to irregular time-series data, aligning more closely with the actual disclosure practices of bond issuers and significantly reducing data completeness requirements. The loss function comprises three components: cross-entropy loss, distribution loss, and separation loss, combining the dual objectives of predictive performance and interpretability.

 Different from the black box characteristics of general machine learning methods, our EMDLOT has the interpretability  capacity and can find several reasons why bonds fail. One is that  our interpretability analysis reveals defaults often stem from liquidity shocks, with firms reliant on external financing struggling near maturity \citep{zhou2023bond}. Another is that the default also stems from long-term structural weaknesses like high leverage erode solvency over time \citep{segal2023overview, wang2023financial}. The default is also relative to the  prospectus risk sections which can highlight pre-issuance vulnerabilities, amplified by incomplete disclosures \citep{yao2024impact, li2023does}. The interpretability of  EMDLOT also shows that, at the macro level, the default of debt can be relative to the weaker GDP growth and adverse trade conditions which can    strain debt capacity \citep{giesecke2011corporate, atawnah2023does}.

Summarily, Our work has several marginal   contributions.

\begin{itemize}
  \item   We  propose a pioneering integration of bond prospectuses into multimodal bond Default prediction.
Existing multimodal bond default prediction studies typically combine numerical financial data with textual sources like news headlines \citep{tang2024unlocking}, earnings call transcripts \citep{tavakoli2025multi}, or credit rating reports \citep{lu2024corporate}. However, these sources have critical flaws: news lacks legal binding and depth, earnings calls are biased toward short-term sentiment, and credit reports may suffer from endogeneity. This paper addresses this gap by introducing bond prospectuses-legally required pre-issuance disclosure documents-as a core textual modality, alongside corporate financial and macroeconomic indicators.

Bond prospectuses offer unique value: their "Risk Factor" and "Important Notice" sections provide objective, pre-issuance risk signals directly linked to default. \citet{yao2024impact} confirm prospectus risk disclosures shape credit premiums, while \citet{li2023does} show low-quality prospectuses correlate with higher defaults. The paper processes prospectuses via GLM-4-9B (key information extraction) and Chinese-BERT-wwm ({embedding}), avoiding high-dimensional sparsity plaguing frequency-based text models \citep{tan2023survey}. Unlike \citet{tang2024unlocking}'s news-based text analysis or \citet{lu2024corporate}'s credit report focus, prospectuses reduce bias and capture granular pre-issuance vulnerabilities. Ablation tests show removing textual data (including prospectuses) drops Recall by 12.71\% and F1-score by 13.63\%, proving their irreplaceable role in multimodal fusion.
  \item  Our work has the clear methodological innovation for irregular time-series and intrinsic interpretability.
Irregular financial time-series (e.g., delayed disclosures, missing data) hinder traditional deep learning models-standard LSTM \citep{ren2023study} and CNN \citep{meng2024novel} require fixed intervals, distorting data. This paper adapts Time-Aware LSTM (T-LSTM) \citep{baytas2017patient} to financial data, using time intervals to decay outdated memory states. This lets the model handle irregularities (e.g., a firm's delayed quarterly report) more accurately than standard LSTM, which treats all time steps equally.

To boost interpretability (a key financial requirement ignored by most deep learning models like ANN and standard LSTM), the paper adds soft clustering and multi-level attention. {Soft clustering groups heterogeneous samples into probabilistic clusters} \citep{aguiar2022learning}, capturing complex risk profiles (e.g., a firm with both liquidity and leverage risks). Multi-level attention (chapter, feature, modality) highlights critical signals: "Risk Factor" sections get 59.17\% textual attention, numerical data 78.10\% modal attention, and cash flow indicators top priority pre-default. Unlike post-hoc methods  {such as LIME/SHAP} \citep{mohsin2025explaining}, these mechanisms are intrinsic. Ablation tests show removing attention cuts Recall by 40.73\%, and removing clustering drops Recall by 23.51\%, confirming their role in balancing accuracy and interpretability-EMDLOT's Recall (0.7547) outperforms LSTM (0.6703) and XGBoost (0.5806) while remaining interpretable.

  \item
 We establish a multi-class classification framework for granular default prediction.
Virtually all prior studies \citep{altman1968financial, bao2023systematic, cui2025bond} frame bond default prediction as binary ("default/non-default"), oversimplifying real-world scenarios where firms often enter intermediate states (e.g.,  {bond} extension) before full default (e.g., Evergrande's 2021 extension before 2022 default \citep{altman2022has}). This paper introduces a three-class framework: "Performing" (normal repayment), "Extended" ({bond} maturity extended), and "Defaulted" (full default).

This design aligns with practical risk management-lenders need different strategies for each class (e.g., monitoring "Extended" firms vs. recovering from "Defaulted" ones). Using SMOTE to balance the imbalanced dataset (1.91\% default rate), EMDLOT outperforms benchmarks in multi-class metrics: its Recall (0.7547) is 8.44 percentage points higher than LSTM (0.6703), and mAP (0.8323) exceeds Random Forest (0.7369). Unlike \citet{tang2024unlocking}'s binary "first-time default" focus or \citet{peng2025forecasting}'s ensemble binary models, this framework captures default progression, enabling proactive intervention. For example, a firm classified as "Extended" can be flagged for restructuring before full default, reducing losses.
\end{itemize}

The remaining sections of this paper include the following: Section~\ref{sec2} presents a review of related literature and the research gaps. Section~\ref{sec3} discusses the research methods and model construction used in this paper. Section~\ref{sec4} introduces the dataset and the data preprocessing procedures. Section~\ref{sec5} outlines the process and results of empirical analysis. Finally, Section~\ref{sec6} presents the main conclusions of this paper.

\section{Literature review}
\label{sec2}

\subsection{Prediction methods}

In bond default prediction, both statistical methods and machine learning methods have been widely applied, each offering distinct advantages in interpretability and predictive performance. Notably, deep learning has emerged as a promising approach, particularly in handling multimodal data integration for improved risk assessment.

Statistical models have long been instrumental in credit risk evaluation. \citet{altman1968financial} introduced the Z-score model based on multiple discriminant analysis, marking a significant advancement in default prediction. \citet{duan2012multiperiod}, who proposed a Forward Intensity Model to estimate corporate default probabilities across different time horizons. While their model demonstrated high accuracy in short-term (three-month) forecasts, predictive performance declined to approximately 60\% as the time horizon extended. More recently, \citet{bao2023systematic} applied a Merton model, akin to the CAPM framework, highlighting the critical role of macroeconomic indicators in bond default risk.

In machine learning, including deep learning, algorithms such as Logistic Regression, Decision Trees, Random Forest, SVM, KNN, XGBoost, and LightGBM have been extensively employed in building bond default prediction models. \citet{cui2025bond} improved upon the traditional Altman Z-score by integrating Ridge Logistic Regression, Support Vector Classifier, and Random Forest, achieving superior predictive accuracy. \citet{peng2025forecasting} integrated Random Forest, GBM, LightGBM, Logistic Regression, SVM, XGBoost, and CNN, aggregating the prediction results of these models into the final prediction through simple voting or stacking methods. They found that LightGBM performed the best, while CNN performed the worst. \citet{heger2024analyzing} systematically compared OLS, LPR, SVM, RF, BART, BRANN, and LSTM, confirming that machine learning models generally surpass traditional linear regression in predictive capability. \citet{meng2024novel} tranformed time-series data into 2D images to relflect their dynamic characteristics, developing a URP-CNN model to access bond credit risk in Chinese listed companies. \citet{ren2023study} combined the characteristics of CNN and LSTM and proposed ConvLSTM for bond default prediction. \citet{tang2024unlocking} employed SVM, XGBoost, and ANN to process numerical data, while using BERTopic for textual data, with a specific focus on predicting the first-time corporate bond defaults, particularly within a 3-month timeframe.

In similar applications for financial risk prediction, \citet{tavakoli2025multi} employed CNN to extract local patterns from numerical data, RNN to process time-series information, and BERT to analyze textual data for bond rating prediction. Similarly, \citet{che2024predicting} utilized BERT for text processing and GRU for time-series analysis to forecast financial distress. \citet{mai2019deep}, in applying CNN to predict bankruptcy, found that the effectiveness of deep learning depends critically on the data modality.

\subsection{The use of multimodality}

In multimodal financial analysis, the primary data modalities include numerical, textual, image, audio, and video modalities. Deep learning techniques have demonstrated remarkable capability in extracting structured patterns from large-scale unstructured multimodal data \citep{dell2025deep}. And new modalities are constantly being explored, such as network modality \citep{che2024predicting}.

Regarding text modality, there are some studies focus on financial statements and credit rating reports \citet{lu2024corporate}, these conventional sources present limitations - financial statements often contain redundant information with quantitative metrics, while credit reports may introduce endogeneity issues or even reverse causality. Researchers also have increasingly explored alternative textual sources such as news headlines \citep{tang2024unlocking}, earnings call transcripts \citep{tavakoli2025multi}, and user-generated texts \citep{kriebel2022credit}.

As legally mandated disclosure documents prepared by issuers before bond offerings, bond prospectuses play a vital role in financial markets: they provide stakeholders with key information for investment decisions and ensure bondholders' right to access material events and risk factors. Related research highlights several critical features of these documents: their risk disclosure sections influence credit risk premiums by shaping investors' risk perception \citep{yao2024impact}. Low-quality prospectuses are linked to higher default probabilities \citep{li2023does}. Furthermore, the greater the discrepancy between the textual tone in bond prospectuses and company's actual financial performance, the wider the credit spread at bond issuance. This discrepancy reduces corporate information transparency, consequently elevating default risk. This aligns with how investors typically cross-validate numerical performance data with textual information to assess risks.
Our deep learning model also follows this multimodal approach.

In natural language processing, word vector technology has developed from static representation to dynamic context modeling. Word2Vec learns fixed word vectors through local context windows and cannot handle polysemous words. BERT, on the other hand, is based on the Transformer architecture and generates dynamic word vectors through a bidirectional self-attention mechanism, which can accurately distinguish word meanings according to context \citep{devlin2019bert}. It performs well in converting text vectors and supporting economic decision-making.
Therefore, we use BERT for relevant text processing. The specific operations will be described in detail later in the text.

\subsection{The use of interpretability}

Traditional post-hoc interpretability frameworks can only serve as supplements and cannot reveal underlying computations or internal representations.

In financial prediction, many scholars are exploring intrinsic interpretability. \citet{tatsat2025beyond} proposed Mechanistic Interpretability, which reverse-engineers models to understand their internal components and identifies the most influential neurons and attention heads in tasks, such as fraud detection and credit scoring. \citet{zhang2025method} argue that intrinsically interpretable models should align with sample features derived from LIME and SHAP methods. They compared Logistic Regression, Decision Trees, Random Forest, and XGBoost, finding that Logistic Regression performed best in intrinsic interpretability, while XGBoost performed the worst. \citet{nazemi2024interpretable} demonstrated that Explainable Boosting Machines (EBM) and Generalized Additive Model Networks (GAMI-Net) can effectively predict corporate bond default recovery rates and directly explain variable impacts without post-hoc methods. \citet{galil2023prediction} used Classification and Regression Trees (CART) to generate if-then-else rules, ensuring intrinsic interpretability and theoretical consistency. \citet{tang2024predicting} compared intrinsic interpretability models: including EBM, GLM, GAM, ReLU-DNN, and FIGS-with post-hoc methods (PFI, SHAP, and PDP) to analyze systemic financial risk drivers. \citet{che2024predicting} incorporated attention mechanisms, quantifying feature importance via average attention weights.

Other disciplines offer similar insights. \citet{aguiar2022learning} used LSTM models with Electronic Health Records (EHR) data to predict patient outcomes, clustering patients and analyzing attention maps to identify critical clinical features and time points. For multimodal models, \citet{wenderoth2025measuring} proposed InterSHAP, which interprets predictions through cross-modal interactions.

\subsection{Research gap}

Bond default prediction has sparked extensive discussions in academic circles, and Table~\ref{tab1} compares some representative literature. From a data perspective, first, time-series features have been widely captured, with only a small number of studies still using static indicators. Second, the use of multimodal data for bond default prediction has only gradually gained attention in recent years. From the perspective of intrinsic interpretability, the research timeline clearly shows a progression: from interpretable linear models to non-interpretable black-box models, and then to the process of "opening the black box." From the perspective of classification objectives, all the listed literature defines bond default prediction as a binary classification task. Our work is the first to simultaneously focus on time-series, multimodality, intrinsic interpretability, and multi-classification.

\begin{table}[htbp]
\centering
\caption{Comparison of related studies}
\label{tab1}
\begin{tabular}{l *{4}{>{\centering\arraybackslash}m{0.15\linewidth}}}
\toprule
Study &
\centering\arraybackslash\makecell{Use of\\time-series} &
\centering\arraybackslash\makecell{Use of\\multimodal} &
\centering\arraybackslash\makecell{Use of intrinsic\\interpretability} &
\centering\arraybackslash\makecell{Use of\\multi-classification} \\
\midrule
\citet{altman1968financial}     &         &         & \checkmark &         \\
\citet{duan2012multiperiod}     & \checkmark &         & \checkmark &         \\
\citet{mai2019deep}             & \checkmark & \checkmark &         &         \\
\citet{bao2023systematic}       & \checkmark &         & \checkmark &         \\
\citet{cui2025bond}             & \checkmark &         &         &         \\
\citet{peng2025forecasting}     & \checkmark &         & \checkmark &         \\
\citet{heger2024analyzing}      & \checkmark &         & \checkmark &         \\
\citet{meng2024novel}           & \checkmark &         &         &         \\
\citet{ren2023study}            & \checkmark &         &         &         \\
\citet{tang2024unlocking}       & \checkmark & \checkmark &         &         \\
\citet{tavakoli2025multi}       & \checkmark & \checkmark &         &         \\
\citet{che2024predicting}       & \checkmark & \checkmark & \checkmark &         \\
\citet{lu2024corporate}         &         & \checkmark &         &         \\
\citet{zhang2025method}         &         &         & \checkmark &         \\
\citet{nazemi2024interpretable} & \checkmark &         & \checkmark &         \\
Our work                        & \checkmark & \checkmark & \checkmark & \checkmark \\
\bottomrule
\end{tabular}
\end{table}

\section{Method}
\label{sec3}

\begin{figure}[htbp]
    \centering
    \includegraphics[width=\textwidth]{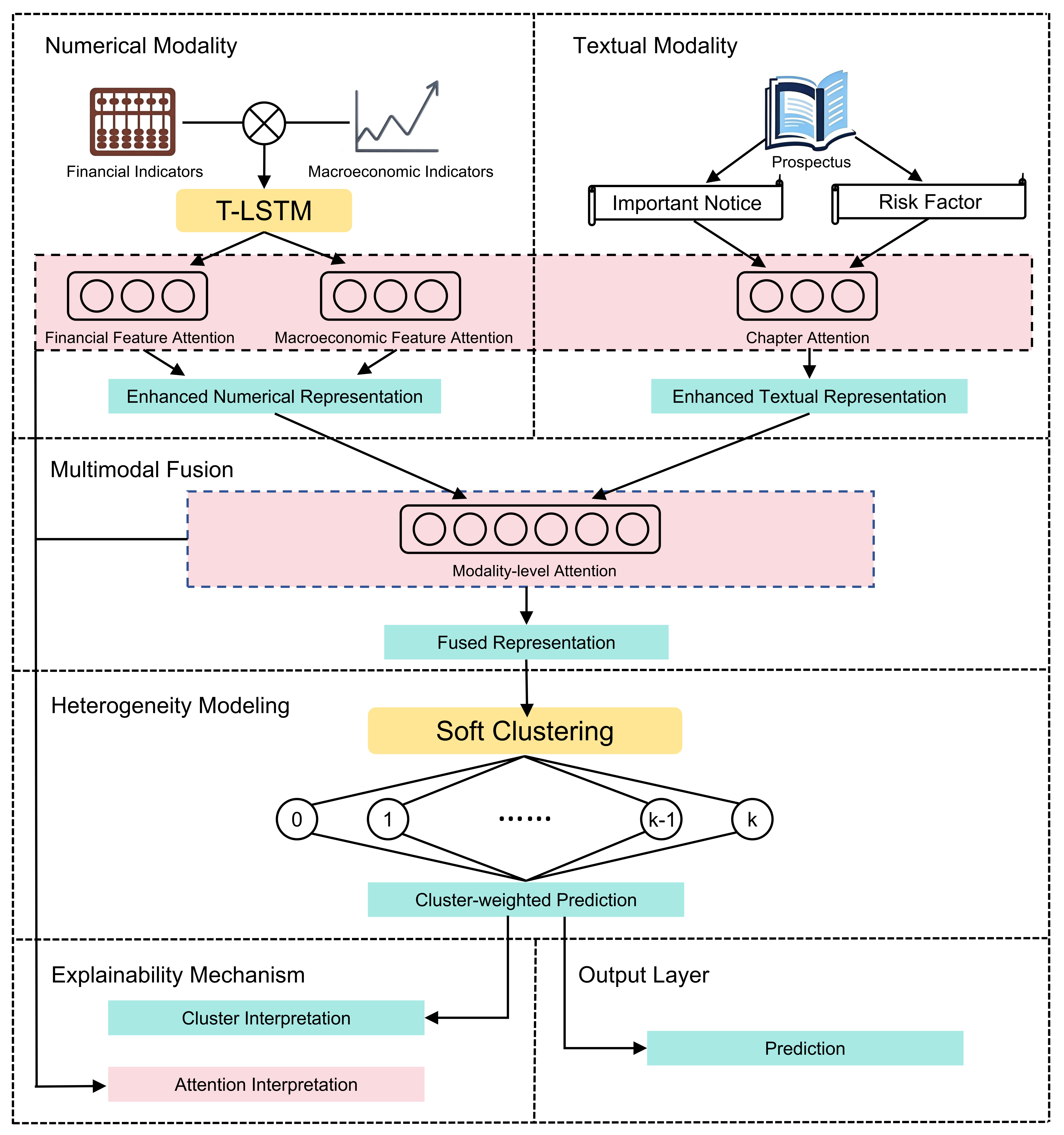}
    \caption{Framework of EMDLOT}
    \label{fig1}
\end{figure}

To harness the great potential of unstructured data in improving bond default prediction, while addressing the substantial challenges in achieving intrinsic interpretability, we propose EMDLOT (Explainable Multimodal Deep Learning fOr Time-series), a unified architecture that integrates structured financial time series with unstructured textual disclosures in an interpretable deep learning framework. Figure~\ref{fig1} illustrates the framework of EMDLOT.

\subsection{T-LSTM for irregularity}

The Time-aware Long Short-Term Memory (T-LSTM) model was first introduced by \citet{baytas2017patient} for clinical time series analysis like electronic health records (EHRs), T-LSTM addresses the limitations of standard LSTM by incorporating elapsed time intervals into its memory update mechanism. Specifically, it decays the influence of prior memory states based on the time gap between observations, thereby effectively reducing the impact of outdated information on current outputs. This temporal decay mechanism makes T-LSTM particularly effective and advantageous in handling sequential data with irregular time gaps.

Financial data of \textcolor{red}{bond}-issuing entities often exhibits irregular time intervals due to varying publication frequencies, missing data, or event-driven disclosures, which fit perfectly with the application of the T-LSTM that introduces a decay gate to modulate long-term memory contributions based on the time elapsed $\Delta_t$ between observations.

At each time step $t$, the decayed memory state $\widetilde{c}_{t-1}$ is computed as:
\begin{align}
    T_t = \sigma(\mathbf{W}_T \Delta_t + \mathbf{b}_T) \\
    \widetilde{c}_{t-1} = T_t \odot c_{t-1}
\end{align}

The standard LSTM update equations are then modified as follows:
\begin{align}
    i_t = \sigma(\mathbf{W}_i x_t + \mathbf{U}_i h_{t-1}) \\
    f_t = \sigma(\mathbf{W}_f x_t + \mathbf{U}_f h_{t-1}) \\
    o_t = \sigma(\mathbf{W}_o x_t + \mathbf{U}_o h_{t-1}) \\
    \widetilde{c}_t = \tanh(\mathbf{W}_c x_t + \mathbf{U}_c h_{t-1}) \\
    c_t = f_t \odot \widetilde{c}_{t-1} + i_t \odot \widetilde{c}_t \\
    h_t = o_t \odot \tanh(c_t)
\end{align}

The variable $\Delta_t \in \mathbb{R}$ denotes the elapsed time between the current observation at step $t$ and the previous valid observation at $t-1$, typically measured in months. This time gap is essential for capturing the irregular nature of financial indicators. To model the influence of this elapsed time, a learnable decay gate is introduced via the transformation $T_t = \sigma(\mathbf{W}_T \Delta_t + \mathbf{b}_T)$, where $\mathbf{W}_T \in \mathbb{R}^{h \times 1}$ and $\mathbf{b}_T \in \mathbb{R}^h$ are parameters learned during training. The resulting gate $T_t \in (0,1)^h$ serves as a vectorized attenuation factor, controlling how much of the previous memory cell should be retained.

The memory cell from the previous time step, denoted as $c_{t-1} \in \mathbb{R}^h$, encodes long-term state information. This memory is downweighted based on temporal distance, yielding the decayed memory $\widetilde{c}_{t-1} = T_t \odot c_{t-1}$, where $\odot$ denotes element-wise multiplication. The current input vector $x_t \in \mathbb{R}^d$, which includes financial indicators and macroeconomic indicators, and the previous hidden state $h_{t-1} \in \mathbb{R}^h$ are then used to compute the standard LSTM gates. Specifically, the input-to-hidden weight matrices $\mathbf{W}_i, \mathbf{W}_f, \mathbf{W}_o, \mathbf{W}_c \in \mathbb{R}^{h \times d}$ and the hidden-to-hidden recurrent matrices $\mathbf{U}_i, \mathbf{U}_f, \mathbf{U}_o, \mathbf{U}_c \in \mathbb{R}^{h \times h}$ are used to compute the input gate $i_t$, forget gate $f_t$, and output gate $o_t$, all of which lie in the range $(0,1)^h$.

The candidate cell update $\widetilde{c}_t$ is computed as a non-linear transformation of the current input and hidden state, i.e., $\widetilde{c}_t = \tanh(\mathbf{W}_c x_t + \mathbf{U}_c h_{t-1})$. The final memory cell is updated by combining the decayed previous memory and the new candidate via the weighted sum $c_t = f_t \odot \widetilde{c}_{t-1} + i_t \odot \widetilde{c}_t$. Lastly, the updated hidden state is computed as $h_t = o_t \odot \tanh(c_t)$, which is passed forward either to the next time step in the sequence or to downstream modules.

This mechanism enables the model to effectively discount outdated financial information and adaptively focus on temporally relevant signals, which is essential for robust modeling of irregular sequences in corporate financial data.

\subsection{Clustering for heterogeneity}

Recognizing that firms of different types are susceptible to varying sources of default risk, we introduce a soft clustering module atop the fused multimodal representations to capture such heterogeneity. This module enhances the model's robustness and provides interpretable insights into distinct risk pathways across issuer subpopulations.

We let $\mathbf{z}_\text{fusion} \in \mathbb{R}^d$ denote the combined embedding obtained after applying the attention-based multimodal fusion described in Section 3.3. This fused vector is passed through a cluster identification network $f_\text{cluster}(\cdot)$, typically implemented as a multilayer perceptron (MLP), which projects it into a $K$-dimensional logit space. A softmax transformation is then applied to generate a probabilistic cluster assignment vector $\pi \in \mathbb{R}^K$,
such that each component $\pi_k$ reflects the degree to which the instance belongs to cluster $k$ and $\sum_{k=1}^K \pi_k = 1$ holds by construction:
$$
   \pi = \text{softmax}(f_\text{cluster}(\mathbf{z}_\text{fusion}))
$$
Each cluster $c_k$ is associated with a dedicated prediction head $P(c_k) \in \mathbb{R}^C$, where $C$ denotes the number of output classes (e.g., default vs. non-default). These classifiers are independent of each other and are trained jointly within the overall model. The final prediction $\hat{\mathbf{y}}$ is computed as the expectation over cluster-specific predictions, weighted by the soft assignment vector:
\begin{align}
    \hat{\mathbf{y}} = \sum_{k=1}^K \pi_k \cdot P(c_k)
\end{align}
This soft mixture-of-experts formulation enables the model to express different decision pathways conditioned on the latent cluster membership of each input, while remaining fully differentiable and interpretable. The probabilistic assignments also support the design of regularization losses that encourage semantic separation between clusters and promote balanced usage across them.

\subsection{Attention for interpretability}

To provide fine-grained interpretability and emphasize salient signals across the input space, we incorporate attention mechanisms at multiple levels of the model architecture \citep{vaswani2017attention}, including chapter-level attention for textual data, feature-level attention for numerical indicators, and modality-level attention for multimodal fusion.

At the chapter level, textual information from bond prospectus disclosures,  including the Risk Factor and Important Notice sections is first encoded independently into chapter embeddings $\mathbf{h}_j \in \mathbb{R}^d$. These embeddings are then aggregated using a chapter-level attention mechanism that assigns greater weights to more informative chapters. The attention score for each chapter is computed as:
\begin{align}
    \alpha_j = \frac{\exp(\mathbf{w}^\top \tanh(\mathbf{h}_j)/\tau)}{\sum_k \exp(\mathbf{w}^\top \tanh(\mathbf{h}_k)/\tau)}
\end{align}
and the aggregated textual representation is given by:
\begin{align}
    \mathbf{h}_\text{text} = \sum_j \alpha_j \mathbf{h}_j
\end{align}
where $\mathbf{w} \in \mathbb{R}^d$ is a learnable attention projection vector and $\tau \in \mathbb{R}^+$ is a temperature parameter that controls the sharpness of attention distribution.

For the numerical modality, which comprises firm-level financial ratios and macroeconomic indicators, we similarly apply feature-level attention over their projected representations. Given a set of encoded features $\mathbf{f}_i \in \mathbb{R}^d$, the attention mechanism computes the importance of each feature as:
\begin{align}
    \alpha_i = \frac{\exp(\mathbf{w}^\top \tanh(\mathbf{f}_i)/\tau)}{\sum_j \exp(\mathbf{w}^\top \tanh(\mathbf{f}_j)/\tau)}
\end{align}
and the aggregated numerical representation is obtained as:
\begin{align}
    \mathbf{f}_\text{agg} = \sum_i \alpha_i \mathbf{f}_i
\end{align}
This mechanism highlights the most relevant financial indicators for each individual prediction.
To adaptively integrate the textual modality and numerical modality, we introduce a modality-level attention mechanism that dynamically balances their contributions based on context. Each modality (indexed by $m \in \{\text{text}, \text{numeric}\}$) is represented by an embedding $\mathbf{z}_m \in \mathbb{R}^d$, and attention weights are computed as:
\begin{align}
    \beta_m = \frac{\exp(\mathbf{w}_m^\top \mathbf{z}_m / \tau)}{\sum_k \exp(\mathbf{w}_k^\top \mathbf{z}_k / \tau)}
\end{align}
leading to a final fused representation:
\begin{align}
    \mathbf{z}_\text{fusion} = \sum_m \beta_m \mathbf{z}_m
\end{align}
where $\mathbf{w}_m \in \mathbb{R}^d$ are modality-specific projection vectors.

These multi-level attention mechanisms collectively enhance the model's interpretability by selectively emphasizing the most informative textual chapters, numerical features, and modalities, thereby enabling more transparent and context-aware predictions.

\subsection{Loss optimization}

The total loss function is designed to jointly optimize prediction accuracy and embedding space structure by incorporating three components: a standard classification loss, a cluster distribution loss to encourage balanced usage of all clusters, and a cluster separation loss that maximizes the distances between cluster centers. Formally, the overall objective function is expressed as
\begin{align}
    \mathcal{L} = \mathcal{L}_\text{cls} + \lambda_\text{dist} \cdot \mathcal{L}_\text{dist} + \lambda_\text{clus} \cdot \mathcal{L}_\text{clus}
\end{align}
where $\mathcal{L}_\text{cls}$ denotes the standard cross-entropy loss comparing the predicted label $\hat{y}$ with the ground truth $y$, thereby directly optimizing classification performance:
\begin{equation}
\mathcal{L}_\text{cls} = \text{CrossEntropy}(y, \hat{y})
\end{equation}
To prevent the model from collapsing to a small subset of clusters and thereby encourage more uniform cluster utilization, a cluster distribution loss $\mathcal{L}_\text{dist}$ is introduced, defined as a normalized entropy over the average cluster assignment probabilities $\bar{\pi}_k$ across the training batch:
\begin{equation}
\mathcal{L}_\text{dist} = -\frac{1}{\log K} \sum_{k=1}^K \bar{\pi}_k \log \bar{\pi}_k
\end{equation}
Here, $\bar{\pi}_k$ represents the mean probability of assignment to cluster $k$, and this term encourages the cluster usage distribution to approach uniformity. Furthermore, to promote semantic distinctness and prevent cluster centers from collapsing toward each other, a cluster separation loss $\mathcal{L}_\text{clus}$ is imposed, which penalizes the inverse of the average pairwise Euclidean distance among the learnable cluster centroids $\mathbf{c}_i \in \mathbb{R}^d$:
\begin{equation}
\mathcal{L}_\text{clus} = -\frac{2}{K(K-1)} \sum_{i < j} \| \mathbf{c}_i - \mathbf{c}_j \|_2
\end{equation}
This separation loss effectively pushes cluster centers apart in the embedding space, fostering more interpretable and well-structured latent representations. The hyperparameters $\lambda_\text{dist}$ and $\lambda_\text{clus}$ control the relative influence of the distribution and separation losses, balancing the model's focus between classification accuracy and embedding space regularization.

\subsection{Initialization}

To mitigate instability during the early stages of training, the model adopts a cluster initialization strategy based on K-Means. Specifically, prior to end-to-end training, we apply the K-Means algorithm to the initial encoder outputs to identify latent cluster structures. The resulting cluster centroids are then used to initialize the parameters of the cluster centers in the model. To ensure alignment between the learned soft assignments and the initial clustering structure, the cluster identifier network is pre-trained to predict the discrete K-Means labels, effectively serving as a supervised warm-up phase.

\section{Dataset and preprocessing}
\label{sec4}

This study uses quarterly data from a total of 1,994 enterprises that issued bonds on the Shanghai Stock Exchange and Shenzhen Stock Exchange between 2015 and 2024, with a sample span of 10 years. The sample includes 38 enterprises that experienced bond defaults, among which 24 were extended and 14 were defaulted. The overall default rate of the sample is 1.91

\begin{figure}[htbp]
    \centering
    \includegraphics[width=0.75\linewidth]{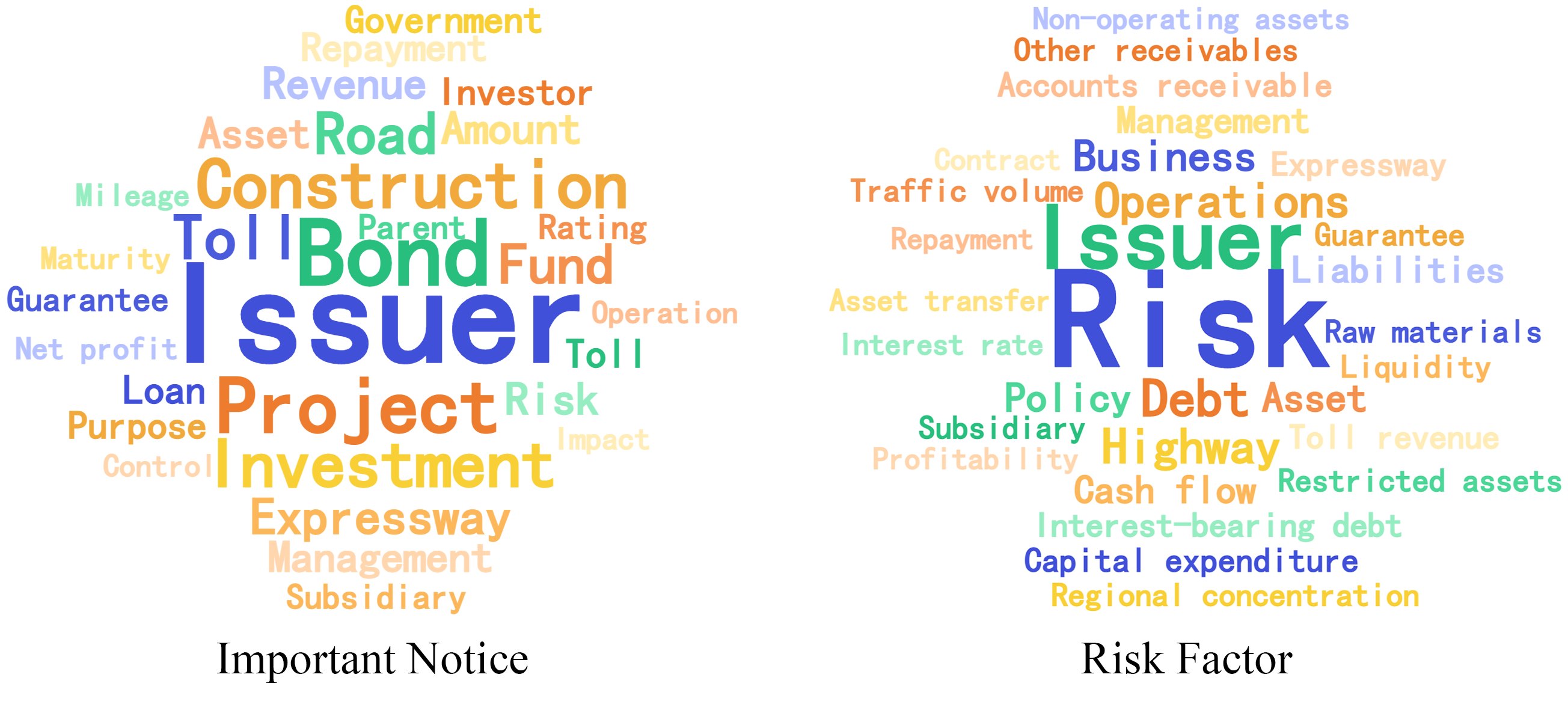}
    \caption{Word clouds of the two sections}
    \label{fig2}
\end{figure}

Textual data is sourced from bond prospectuses disclosed on the official websites of the Shanghai Stock Exchange\footnote{Shanghai Stock Exchange, bond information platform. Retrieved March 20, 2025, from \url{https://bond.sse.com.cn/bridge/home/}} and Shenzhen Stock Exchange\footnote{Shenzhen Stock Exchange, bond announcements. Retrieved March 20, 2025, from \url{https://www.szse.cn/disclosure/bond/notice/index.html}}. We extracted text content from two sections as the objects of analysis: Important Notice and Risk Factor (in some documents, this section is titled Risk Reminder and Explanation). To illustrate the textual characteristics of the two sections, Figure~\ref{fig2} shows word clouds from the Important Notice and Risk Factor sections of the bond prospectuses of Gansu-Highway Traffic Construction Group Co., Ltd.

\begin{figure}[htbp]
    \centering
    \includegraphics[width=0.75\linewidth]{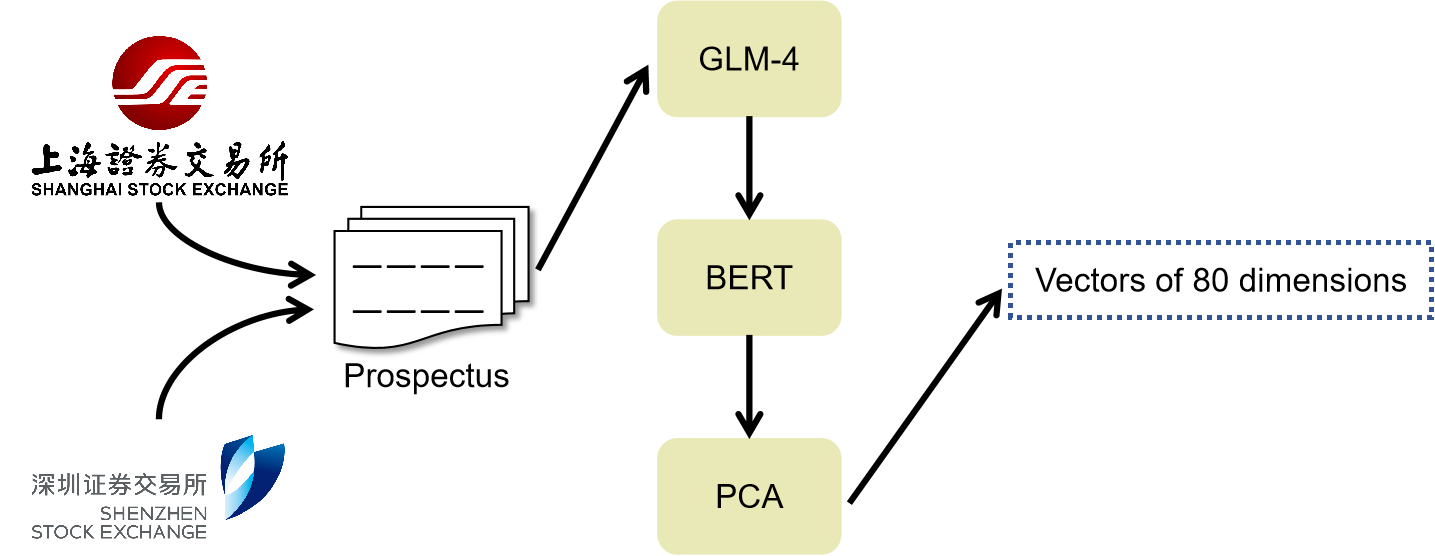}
    \caption{Preprocessing of the original textual data}
    \label{fig3}
\end{figure}

Given that the BERT model has a single input limit of 512 tokens, this paper first uses the GLM-4-9B (THUDM/GLM-4-9B-0414) large model to extract key content from each of the two sections separately. It then uses the Chinese-BERT-wwm model to  {embed} this content into 768-dimensional  {embeddings}. The  {embeddings} of the two sections are concatenated to form a 1536-dimensional text feature vector, which is subsequently reduced to 80 dimensions using Principal Component Analysis (PCA), with 40 dimensions retained for each section. The cumulative explained variance ratio of PCA is 0.7678, indicating that most of the information is preserved. Figure~\ref{fig3} shows this process.

The numerical features used in this study include 32 financial indicators and 36 macroeconomic indicators\footnote{In Online Appendix, Appendix~\ref{seca} introduces the financial and macroeconomic indicators used in the study, including their abbreviations and summary statistics.}, all sourced from the Wind database, with a time span from 2015 to 2024, covering a maximum of 39 quarters. Specific indicators are shown in Table~\ref{tab2} and Table~\ref{tab3}. To address missing values in the data, Multiple Imputation is employed for filling. This approach not only retains data variability but also avoids the problem of variance underestimation, making it particularly suitable for handling complex missing patterns in high-dimensional economic and financial data.

\begin{table}[htbp]
\centering
\caption{Financial indicators by category}
\label{tab2}
\begin{tabular}{@{}>{\centering\arraybackslash}m{3.5cm} >{\raggedright\arraybackslash}m{9.5cm}@{}}
\toprule
Category & \multicolumn{1}{>{\centering\arraybackslash}m{9.5cm}@{}}{Financial indicators} \\  
\midrule
Profitability &
Earnings before interest and taxes (EBIT), EBIT (estimated), earnings before interest, taxes, depreciation and amortization (EBITDA), EBITDA (estimated), main operating revenue, main operating profit, net profit, main operating profit margin (\%), return on total assets (ROA), return on equity (ROE), EBITDA to total revenue ratio, operating cash flow to EBITDA \\
\midrule
Solvency &
Current ratio, quick ratio, cash ratio, monetary funds to total debt, interest coverage ratio, EBITDA to interest-bearing debt ratio \\
\midrule
Operational capabilities &
Growth rate of main operating revenue (\%), inventory turnover ratio, net cash flow from operating activities \\
\midrule
Finance structure &
Debt-to-asset ratio, short-term debt to total debt, interest-bearing debt to total invested capital, net debt, interest-bearing debt, total debt, net assets, monetary assets, total assets, net cash flow from investing activities, net cash flow from financing activities \\
\bottomrule
\end{tabular}
\end{table}

\begin{table}[htbp]
\centering
\caption{Macroeconomic indicators by category}
\label{tab3}
\begin{tabular}{@{}>{\centering\arraybackslash}m{3.5cm} >{\raggedright\arraybackslash}m{9.5cm}@{}}
\toprule
Category & \multicolumn{1}{>{\centering\arraybackslash}m{9.5cm}@{}}{Macroeconomic indicators} \\
\midrule
National accounts &
Gross domestic product (GDP), GDP-primary industry, GDP-secondary industry, GDP-tertiary industry, GDP growth rate (YoY \%), GDP-primary industry (YoY \%), GDP-secondary industry (YoY \%), GDP-tertiary industry (YoY \%), GDP growth rate (QoQ \%) \\
\midrule
Monetary financials &
Broad money (M2), narrow money (M1), cash in circulation (M0), M2 (YoY \%), M1 (YoY \%), M0 (YoY \%), official foreign exchange reserves, end-of-period exchange rate, total sources of funds of financial institutions, total sources of funds of financial institutions\&financial bonds, aggregate financing to the real economy (AFRE), AFRE (YoY \%), AFRE (QoQ \%) \\
\midrule
Prices &
Consumer price index (CPI), CPI-urban, CPI-rural \\
\midrule
Government finance &
Fiscal revenue, fiscal revenue YoY \%, fiscal expenditure, fiscal expenditure YoY \% \\
\midrule
International trade &
Total imports and exports, total imports and exports (YoY \%), trade balance \\
\midrule
Sentiment &
Business climate index, industrial climate index, economist confidence index, consumer confidence index \\
\bottomrule
\end{tabular}
\end{table}

To more intuitively capture the temporal evolution characteristics of key variables in the data, this paper selects several representative financial indicators and macroeconomic indicators. Their quarterly average values are plotted into trend charts spanning 30 quarters, aiming to illustrate their dynamic structures and potential cyclical features.

\begin{figure}[htbp]
    \centering
    \includegraphics[width=1\linewidth]{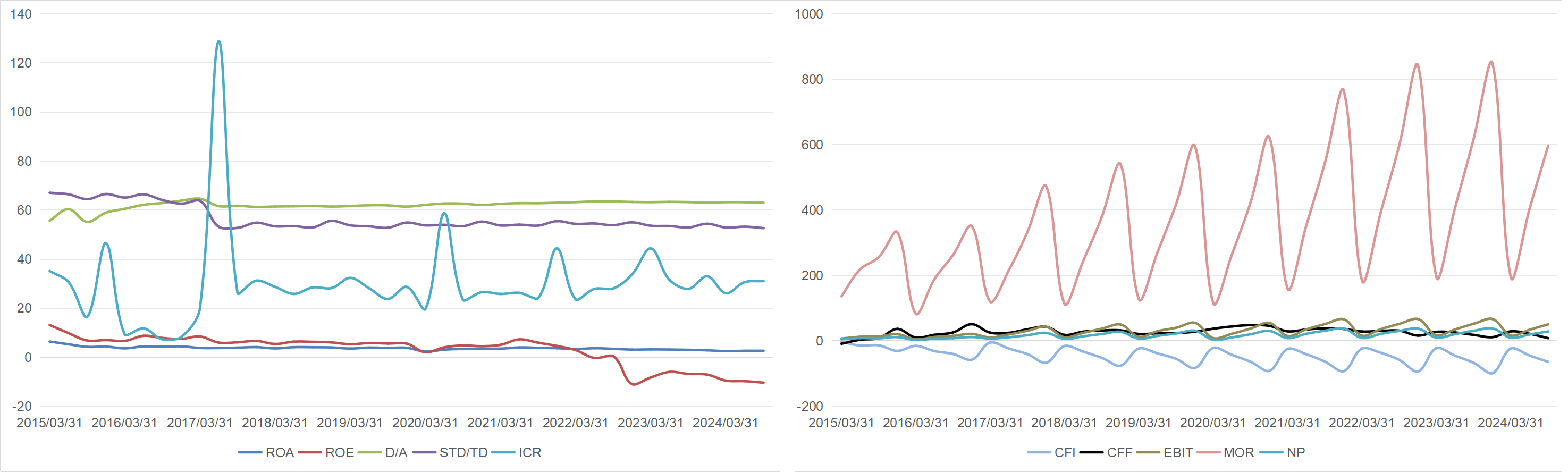}
    \caption{Trends of typical financial indicators}
    \label{fig4}
\end{figure}

Figure~\ref{fig4} shows the quarterly trends of typical financial indicators, including ROA, ROE, Debt-to-Asset Ratio, Short-term Debt to Total Debt, Interest Coverage Ratio, Net Cash Flow from Investing Activities, Net Cash Flow from Financing Activities, EBIT, Main Operating Revenue, and Net Profit. It can be observed that cash flow indicators and some profitability-related indicators exhibit strong cyclicality, which may be influenced by macroeconomic cycle fluctuations or accounting policies (such as revenue recognition rules). In contrast, financial structure ratios exhibit relatively stable overall changes without pronounced cyclicality, as the numerator and denominator may vary in the same direction.

\begin{figure}[htbp]
    \centering
    \includegraphics[width=1\linewidth]{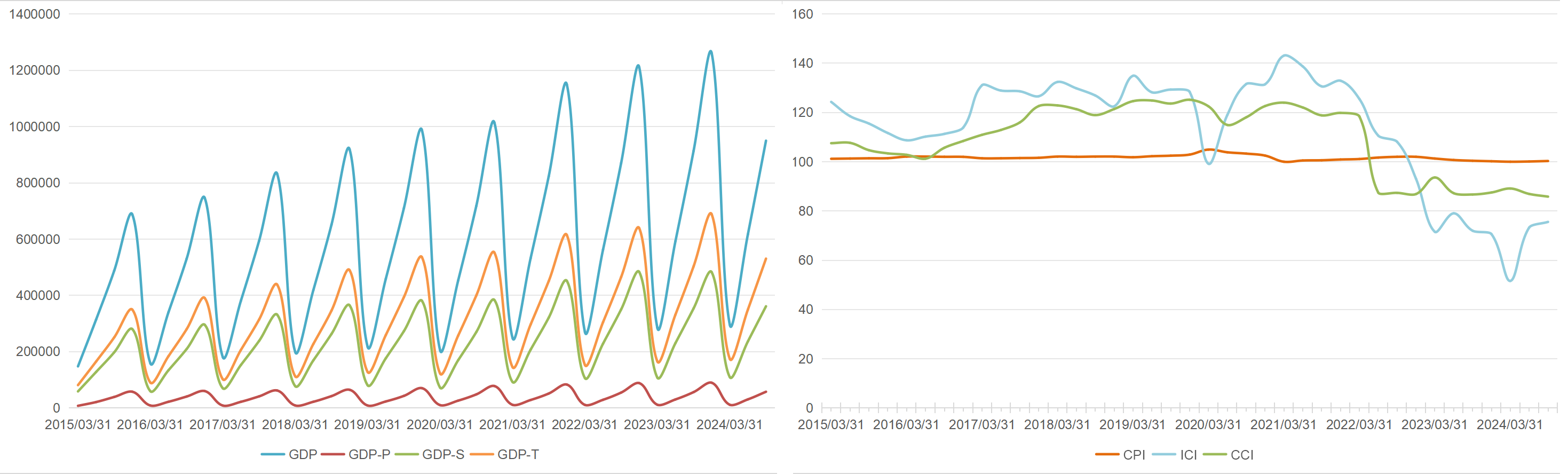}
    \caption{Trends of typical macroeconomic indicators}
    \label{fig5}
\end{figure}

Figure~\ref{fig5} presents the quarterly trends of typical macroeconomic indicators, including GDP, GDP-Primary Industry, GDP-Secondary Industry, GDP-Tertiary Industry, Consumer Price Index, Industrial Climate Index, and Consumer Confidence Index. It can be seen that GDP-related indicators exhibit strong quarterly cyclicality, which may be closely linked to the quarterly implementation of China's fiscal and monetary policies. In contrast, price indices and sentiment indices do not show a unified cyclical pattern. Their fluctuations more reflect non-structural factors such as supply-demand relationships and market expectations, while also being affected by the smoothing of short-term disturbances through statistical methods themselves (e.g., chain Laspeyres method).

To ensure the predictive model maintains forward-looking applicability in real-world scenarios, 
we exclude all financial and macroeconomic indicator data of defaulting enterprises for the two quarters prior to their default occurrence. The purpose of this processing is to simulate the practical scenario of advance prediction of future defaults, preventing the model from utilizing abnormal data that is too close to the default node and thereby avoiding ``post-hoc judgment'' bias in forecasting future trends.

The entire dataset is divided into an 80\% training set and a 20\% testing set. Due to the extremely low proportion of default samples, there is a serious class imbalance issue. In the training phase, this paper employs SMOTE (Synthetic Minority Oversampling Technique) to oversample default samples. The quantity ratio of Performing samples, Extended samples, and Defaulted samples is balanced to 1:1:1, aiming to improve the model's ability to identify minority classes (Extended samples and Defaulted samples). The number of samples in each class before and after processing is shown in Table~\ref{tab4}.

\begin{table}[htbp]
\centering
\caption{Training set sample count comparison}
\label{tab4}
\begin{tabular}{@{}lcc@{}}
\toprule
Category & Before SMOTE & After SMOTE \\
\midrule
Performing & 1565 & 1565 \\
Extended   &   19 & 1565 \\
Defaulted  &   11 & 1565 \\
\midrule
Total & 1595 & 4695 \\
\bottomrule
\end{tabular}
\end{table}

To reflect real-world scenarios, the test set was not processed with SMOTE, and its sample size remains unchanged. During the training and prediction phases, we input both synthetic samples and original training data into the model for training, and then use the trained model for testing.

\section{Empirical analysis}
\label{sec5}

\subsection{Evaluation metrics}

To evaluate the performance of the multi-class bond default prediction model, we adopt four widely used metrics: AUC, Recall, F1-score, and mean Average Precision (mAP)\footnote{In Online Appendix, Appendix~\ref{secb} explains the evaluation metrics-AUC, Recall, F1-score, and mAP-employed to comprehensively assess model performance.}. These indicators together capture the model's ability to discriminate between classes, identify minority categories, and provide reliable probability rankings. Considering the severe imbalance across classes, all metrics are reported using macro averaging, so that each class contributes equally to the overall evaluation.

AUC measures the overall discriminative capacity of the classifier, while Recall emphasizes the detection of default and extended bonds that are often underrepresented in the data. The F1-score, as the harmonic mean of precision and recall, provides a balanced view of accuracy and sensitivity. Finally, mAP evaluates the quality of probability ranking across classes and is particularly suitable for multi-class imbalanced scenarios. Collectively, these metrics provide a comprehensive assessment of model performance from complementary perspectives.

\subsection{Experiment}

During the model training, we introduced the padding + masking mechanism to address the issue of inconsistent sequence lengths. Padding ensures sequences of different lengths are standardized to the same length. Masking is applied to mark the padded segments, ensuring that the model can ignore these non-informative, artificial value during training. This prevents interference with the model's feature learning process from the padded data. Meanwhile, to prevent model overfitting and improve training efficiency, we employed early stopping to monitor the training process. The training will terminate early when the performance does not show significant improvement over consecutive training epochs. The optimal value of the early stopping patience parameter is determined through subsequent hyperparameter optimization\footnote{In Online Appendix, Appendix~\ref{secc} provides experimental details, such as implementation environment, random seeds, and hyperparameter tuning procedures.}.

To achieve the best model performance, we first attempted Grid Search for hyperparameter tuning. However, its performance proved unsatisfactory in practice. Therefore, we switched to the Optuna for hyperparameter optimization. The hyperparameter ranges involved in the experiment and their final selected values are as follows in Table~\ref{tab5}:

\begin{table}[htbp]
\centering
\caption{Optimal hyperparameters and search ranges}
\label{tab5}
\begin{tabular}{lccc}
\toprule
Hyperparameter & Range & Optimal hyperparameter \\
\midrule
hidden\_size & 256, 512, 768, 1024 & 512 \\
dropout & 0.1 to 0.7 (step 0.1) & 0.3 \\
learning\_rate & 1e-5 to 1e-2 (log-uniform) & 0.0006175651418191845 \\
weight\_decay & 1e-6 to 1e-3 (log-uniform) & 1.0162398945608497e-06 \\
num\_clusters & 2 to 8 (integer) & 8 \\
cluster\_loss\_weight & 0.0005 to 0.005 (step 0.0005) & 0.004 \\
dist\_loss\_weight & 0.02 to 0.08 (step 0.005) & 0.035 \\
modal\_temperature & 1.0 to 3.0 (step 0.1) & 2.6 \\
numeric\_feature\_temperature & 0.5 to 0.99 (step 0.05) & 0.9 \\
text\_feature\_temperature & 0.5 to 0.99 (step 0.05) & 0.55 \\
patience & 5 to 20 (integer) & 13 \\
batch\_size & 8, 16, 32 & 8 \\
\bottomrule
\end{tabular}
\end{table}

This study defines bond default prediction as a three-class classification task, where the three categories correspond to performing, extended, and defaulted respectively. The core objective is to identify extended and defaulted companies from the entire set of samples. To ensure the reliability of experimental results, all models undergo independent random seed experiments. A single experiment is deemed valid only if the model successfully predicts all three categories in that specific run. After accumulating a total of 10 valid experiments, the mean and standard deviation of each evaluation metric are calculated and adopted as the final results.

To assess the performance advantages of our proposed model, we selected several classical models from both traditional machine learning and deep learning domains as benchmarks. Specifically, the traditional machine learning models include Logistic Regression, SVM, Decision Tree, KNN, Random Forest, LightGBM, and XGBoost. The deep learning models include ANN and LSTM. All benchmarked models were trained by the same training datasets, and adopted the same evaluation criteria and experimental repetition strategy as the proposed model to ensure the fairness of the comparison.

\subsection{Results and discussions}
\subsubsection{Model performance comparison results}

Table~\ref{tab6} presents the four key performance metrics of each model across ten independent experiments, including Recall, F1-score, mAP, and AUC. Results are reported as mean values with standard deviations to ensure robustness and reproducibility. Notably, SVM  failed to successfully identify all three categories of samples even after numerous experiments, yielding no valid experimental results. As a result, it was excluded from the performance comparison in this study\footnote{In Online Appendix,  Appendix~\ref{secd} presents the soft clustering results, highlighting data heterogeneity and default-related distributional patterns across clusters.}.

Comparative analysis reveals that the EMDLOT model exhibits significantly superior performance across all key benchmarked models in three critical metrics: Recall (0.7547), F1-score (0.7734), and mAP (0.8323). This robust performance underscores its advanced modeling capabilities and generalization efficacy in complex multi-class prediction tasks. Particularly crucial is its leading performance in recall metric. As in practical financial risk control scenarios, the paramount goal in this domain is not merely achieving balanced prediction accuracy across all samples, but maximizing the detection of potential bond defaults to enable timely intervention. In other words, compared to a small number of  false positives, it is more important to avoid false negatives-missing truly defaulting entities. As the saying goes, ``Better safe than sorry,'' which lies at the root of the business's high reliance on recall. Therefore, recall can more authentically and directly reflect the application value and practicality of the model in real-world financial operations.

In contrast, traditional models such as logistic regression, decision tree, and KNN show significant deficiencies in recall performance. For example, the recall of the decision tree stands at only 0.4927, barely meeting the basic requirement for high sensitivity in practical business scenarios and thus carrying a notable risk of false negatives. Furthermore, although XGBoost and LightGBM achieved exceptionally high scores in the AUC metric (0.9873 and 0.9862 respectively), AUC is more focused on evaluating overall classification ability and often exhibits a performance bias favoring the majority class. In the context of tasks with severe class imbalance, such as identifying rare high-risk categories (Extended and defaulted samples), a high AUC may be misleading and fail to accurately reflect the model's actual effectiveness in identifying minority classes.

EMDLOT effectively maintains a robust AUC (0.9435) while simultaneously achieving a significant enhancement in the detection of critical minority classes. This achieves a favorable balance between recall capability and overall discriminative power, reflecting the consistency between its advantageous strategy and optimization objectives in high-risk financial prediction tasks.

\begin{table}[htbp]
\centering
\caption{Comparative analysis of EMDLOT with benchmarked models.}
\label{tab6}
\begin{tabular}{lcccc}
\toprule
Metrics & Recall & F1-score & mAP & AUC \\
\midrule
Logistic Regression & \makecell{0.5110 \\ (0.0753)} & \makecell{0.5324 \\ (0.0654)} & \makecell{0.6040 \\ (0.0784)} & \makecell{0.9346 \\ (0.0361)} \\
Decision Tree & \makecell{0.4927 \\ (0.0882)} & \makecell{0.5102 \\ (0.1162)} & \makecell{0.4379 \\ (0.0943)} & \makecell{0.6575 \\ (0.0667)} \\
KNN & \makecell{0.5413 \\ (0.1147)} & \makecell{0.3564 \\ (0.0270)} & \makecell{0.3587 \\ (0.0118)} & \makecell{0.7493 \\ (0.0549)} \\
Random Forest & \makecell{0.6261 \\ (0.0743)} & \makecell{0.6984 \\ (0.0791)} & \makecell{0.7369 \\ (0.1118)} & \makecell{0.9805 \\ (0.0211)} \\
LightGBM & \makecell{0.5720 \\ (0.1069)} & \makecell{0.6085 \\ (0.1067)} & \makecell{0.7175 \\ (0.1275)} & \makecell{0.9862 \\ (0.0094)} \\
XGBoost & \makecell{0.5806 \\ (0.0884)} & \makecell{0.6010 \\ (0.1037)} & \makecell{0.6851 \\ (0.1111)} & \makecell{\textbf{0.9873} \\ \textbf{(0.0080)}} \\
ANN & \makecell{0.5408 \\ (0.0678)} & \makecell{0.5691 \\ (0.0869)} & \makecell{0.5952 \\ (0.0645)} & \makecell{0.9602 \\ (0.0118)} \\
LSTM & \makecell{0.6703 \\ (0.1233)} & \makecell{0.6893 \\ (0.1035)} & \makecell{0.8006 \\ (0.0869)} & \makecell{0.9422 \\ (0.0723)} \\
EMDLOT & \makecell{\textbf{0.7547} \\ \textbf{(0.0670)}} & \makecell{\textbf{0.7734} \\ \textbf{(0.0521)}} & \makecell{\textbf{0.8323} \\ \textbf{(0.0462)}} & \makecell{0.9435 \\ (0.0625)} \\
\bottomrule
\end{tabular}
\end{table}

Additionally, we ranked the different metrics of each model in Table~\ref{tab7}. It was found that EMDLOT ranked first in all three metrics of Recall, F1-score, and mAP, with an average ranking of 2.0, which is significantly superior to other models. This not only indicates that EMDLOT has achieved leading performance across multiple dimensions but also verifies its consistency and stability in cross-dimensional performance. In particular, it demonstrates stronger robustness when dealing with imbalanced samples and identifying critical minority classes.

While some deep learning models such as LSTM approach EMDLOT in terms of mAP (0.8006), they fall slightly short in Recall and F1-score, with more significant fluctuations in their rankings. This indicates a certain degree of instability and risk in their performance for real-world tasks. Therefore, in high-sensitivity, high-stakes financial default prediction scenarios, EMDLOT offers a more practical and reliable solution.

\begin{table}[htbp]
\centering
\caption{Ranking comparison of EMDLOT against benchmarked models}
\label{tab7}
\begin{tabular}{lccccc}
\toprule
Metrics & Recall & F1-score & mAP & AUC & Avg.\ rank\\
\midrule
Logistic Regression & 8 & 7 & 6 & 7 & 7.0 \\
Decision Tree & 9 & 8 & 8 & 9 & 8.5 \\
KNN & 6 & 9 & 9 & 8 & 8.0 \\
Random Forest & 3 & 2 & 3 & 3 & 2.8 \\
LightGBM & 5 & 4 & 4 & 2 & 3.8 \\
XGBoost & 4 & 5 & 5 & \textbf{1} & 3.8 \\
ANN & 7 & 6 & 7 & 4 & 6.0 \\
LSTM & 2 & 3 & 2 & 6 & 3.3 \\
EMDLOT & \textbf{1} & \textbf{1} & \textbf{1} & 5 & \textbf{2.0} \\
\bottomrule
\end{tabular}
\end{table}

In summary, EMDLOT not only demonstrates significant performance advantages at the metric level but also aligns highly with the financial risk management system in terms of business goal orientation. Its outstanding performance in recall provides solid support for building a default detection system characterized by high recall, strong stability, and low missed detection rates.

\subsubsection{Ablation experiment}

\begin{figure}[htbp]
    \centering
    \includegraphics[width=0.75\linewidth]{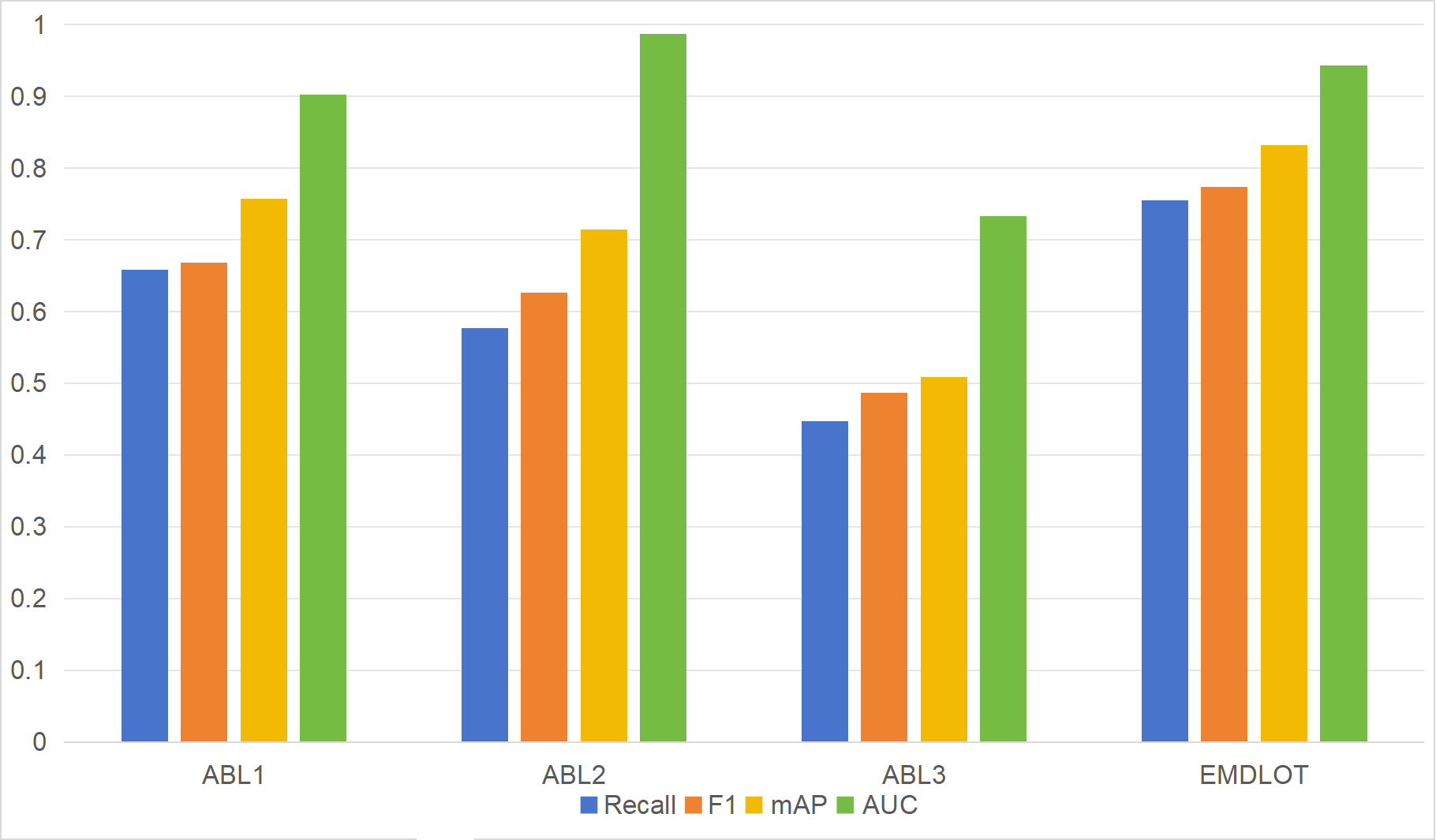}
    \caption{Ablation experiment comparison}
    \label{fig6}
\end{figure}

To verify the effectiveness of each component in the proposed model, we designed and conducted systematic ablation experiments following the principle of Occam's Razor. Specifically, we removed the textual modality (ABL1), soft clustering mechanism (ABL2), and attention mechanism (ABL3) from the model respectively to evaluate the impact of each component on the overall performance. These ablated variants were then compared with the complete model (EMDLOT), and the experimental results are presented in the Table~\ref{tab8} and Figure~\ref{fig6}.

\begin{table}[htbp]
\centering
\caption{Ablation experiment results}
\label{tab8}
\begin{tabular}{lcccc}
\toprule
Metrics & Recall & F1-score & mAP & AUC \\
\midrule
ABL1 & \makecell{0.6588\\(0.1206)} & \makecell{0.6680\\(0.1099)} & \makecell{0.7573\\(0.1048)} & \makecell{0.9029\\(0.0920)} \\
ABL2 & \makecell{0.5773\\(0.0629)} & \makecell{0.6265\\(0.0767)} & \makecell{0.7141\\(0.0850)} & \makecell{\textbf{0.9867}\\\textbf{(0.0091)}} \\
ABL3 & \makecell{0.4473\\(0.0860)} & \makecell{0.4873\\(0.1019)} & \makecell{0.5089\\(0.0751)} & \makecell{0.7336\\(0.1654)} \\
EMDLOT & \makecell{\textbf{0.7547}\\\textbf{(0.0670)}} & \makecell{\textbf{0.7734}\\\textbf{(0.0521)}} & \makecell{\textbf{0.8323}\\\textbf{(0.0462)}} & \makecell{0.9435\\(0.0625)} \\
\bottomrule
\end{tabular}
\end{table}

After removing the text modality, the overall performance of the model also declined, with Recall, F1-score, mAP, and AUC decreasing by 12.71\%, 13.63\%, 9.01\%, and 4.30\% respectively. Although this impact is less significant than that of the aforementioned two modules, it still reflects the value of textual data in multimodal fusion. Meanwhile, we further tested a model retaining only the textual modality, which failed to operate effectively in the three-classification task and the results did not converge. This indicates that the numerical modality plays a crucial role in the current task and is a decisive factor for the model to successfully complete the classification task.

Interestingly, when the soft clustering mechanism was removed, the AUC increased by 4.58\% instead. However, this enhancement was accompanied by a significant degradation in the model's capability to handle heterogeneous data. Specifically, Recall and mAP decreased by 23.51\% and 14.20\% respectively, which confirms that the soft clustering mechanism makes a significant contribution to enhancing the model's adaptability to sample differences.

The removal of the attention mechanism had the most significant impact on model performance. Specifically, Recall, F1-score, mAP, and AUC dropped by 40.73\%, 36.99\%, 38.86\%, and 22.25\% respectively. This indicates that the attention mechanism plays an irreplaceable role in improving the model's ability to capture key information from multimodal data.

In summary, the three ablation experiments all demonstrate that each module contributes to improving model performance to varying degrees. Among them, the attention mechanism makes the most prominent contribution, followed by the soft clustering mechanism and the textual modality. These results verify the rationality and necessity of our model design.

\subsubsection{Interpretability analysis}

We conducted interpretability analysis from two perspectives: static attention mechanisms and dynamic temporal attention. Clustering analysis was performed based on differences in enterprises' financial indicators and text semantic representations, while macroeconomic indicators were excluded because, within any given time window, these variables exhibit negligible inter-enterprise variation. Consequently, macroeconomic attention patterns are visualized via population-level heatmaps rather than cluster-specific mappings. For textual modality, we employed section-level static attention analysis, omitting temporal heatmaps due to its inherent non-sequential nature and absence of cross-quarterly evolutionary characteristics.

\begin{figure}[htbp]
    \centering
    \includegraphics[width=0.5\linewidth]{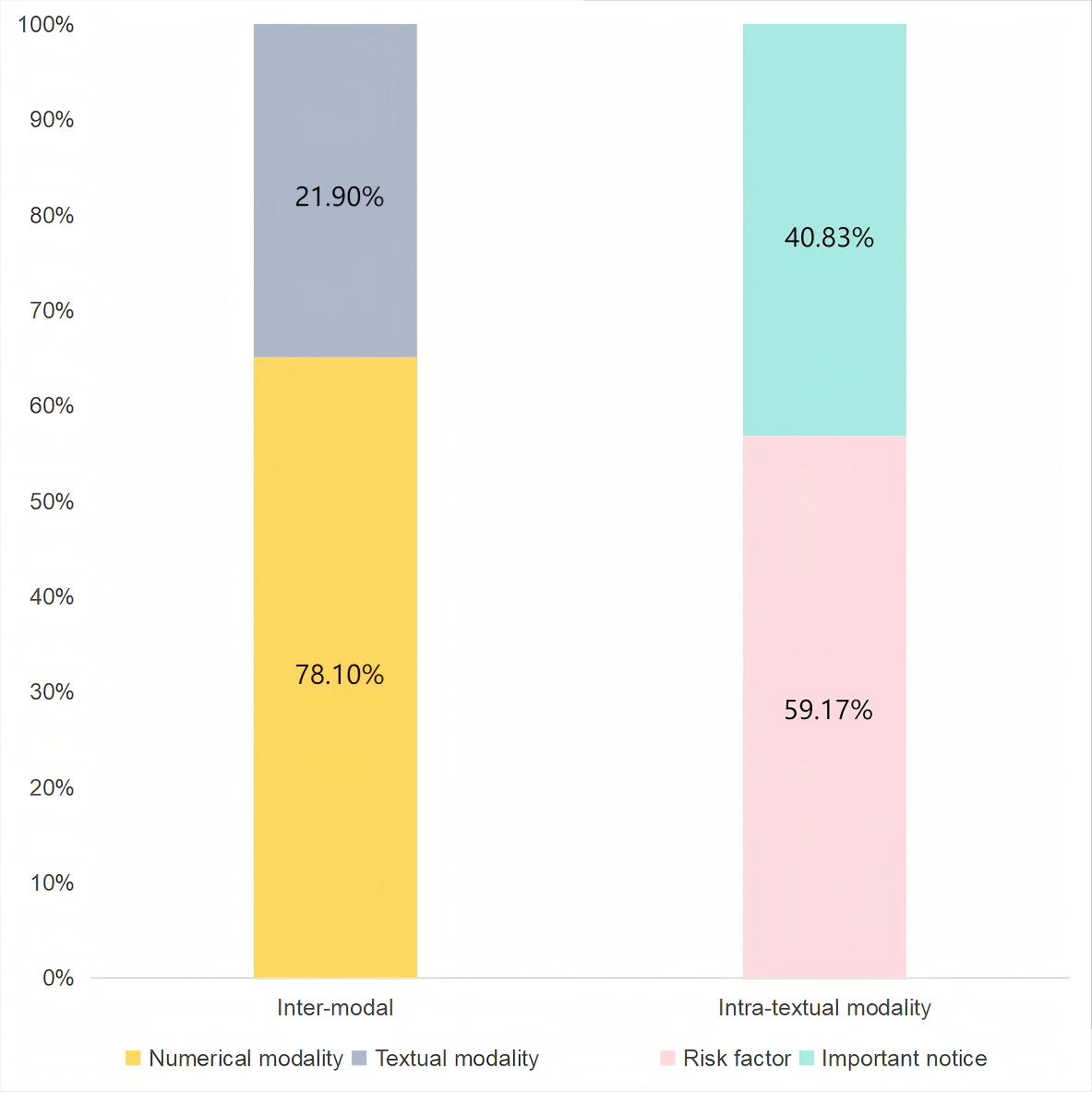}
    \caption{Inter-modal and intra-textual modality attention weight}
    \label{fig7}
\end{figure}

In order to reveal the information from different modalities, we analyzed the static attention structures within the model, including inter-modal attention and intra-textual modality attention at the chapter level. The bar charts in Figure~\ref{fig7} illustrate the average attention distribution of the model at these two levels across ten experimental runs.

In terms of inter-modal attention, the model strongly favors the numerical modality (78.10\%) over the textual one (21.90\%), consistent with ablation results showing that removing numerical data leads to prediction failure and the findings of \citet{stevenson2021value} on Micro, Small, and Medium Enterprises (mSMEs). This highlights the dominant role of structured financial information in default prediction. At the intra-textual level, the model assigns higher attention to the Risk factor chapter (59.17\%) than to the Important Notice (40.83\%), indicating that disclosures of potential risks are particularly critical for identifying defaults. Overall, static attention analysis reveals a hierarchical fusion strategy in which numerical features dominate, textual data supplement, and risk-related content is prioritized.

\begin{figure}[htbp]
    \centering
    \includegraphics[width=1\linewidth]{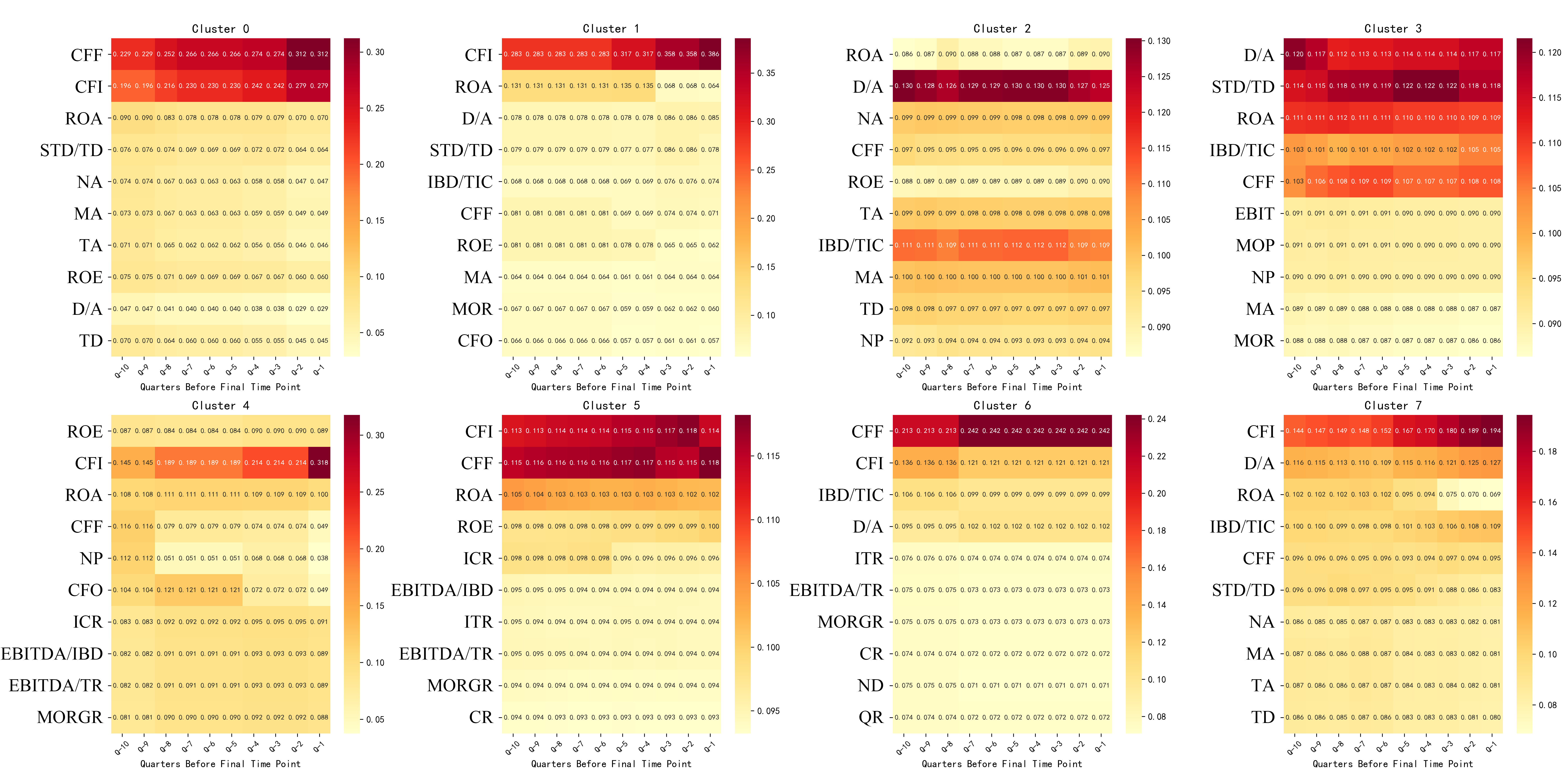}
    \caption{Financial indicators attention map}
    \label{fig8}
\end{figure}

Figure~\ref{fig8} presents attention-based heatmaps of financial features from a representative experimental trial, visualizing the relative importance assigned by the model to distinct financial indicators across eight preceding quarters (Q-8 to Q-1) during default risk prediction. Samples were partitioned into eight clusters (Cluster 0-7), with each subplot depicting the temporal evolution of attention weights for the top 10 financial features within a cluster. Comparative analysis reveals substantial inter-cluster divergence in feature prioritization, highlighting the dynamic time-dependent valuation of financial characteristics across heterogeneous enterprise cohorts. This experiment exhibits representative patterns consistent with majority results.

In several clusters (e.g., Cluster 0, 1, 5, 6, and 7), both net cash flow from investing activities and net cash flow from financing activities consistently receive substantial attention weights across most time points, peaking in the final quarter (Q-1). This pattern suggests that the model relies heavily on firms'  cash flow conditions when assessing default risk, particularly the dynamics of capital expenditures and external financing activities. Such a feature indicates a strong dependence on external funding, underscoring the close linkage between financial stability and financing conditions. Moreover, this reliance may rapidly erode liquidity and intensify short-term debt-servicing pressure. Since Q-1 typically coincides with a critical period preceding repayment obligations, firms that maintain high levels of investment spending while facing restricted access to financing are especially vulnerable to liquidity shortfalls, which can ultimately trigger default.

By contrast, in Clusters 2 and 3, the model places greater emphasis on conventional indicators of financial soundness, such as the debt-to-asset ratio and return on total assets. The temporal distribution of these features is relatively stable, suggesting that in these groups of firms, the model relies more on long-term assessments of leverage and profitability to detect default risk \citep{segal2023overview}. This implies that defaults in these firms may stem primarily from financial mismatches caused by prolonged ineffective investments \citep{wang2023financial}, rather than from sudden liquidity shocks.

Notably, in Cluster 4 the model assigns simultaneously high attention weights to both cash flow-related and capital structure-related indicators, indicating that firms in this cluster require a joint evaluation of liquidity conditions and balance sheet structure when assessing credit risk.

In addition, solvency indicators such as short-term debt to total debt and interest-bearing debt to total invested capital also receive relatively high attention weights across most clusters. The former reflects the share of debt maturing in the short term, capturing liquidity management pressure, while rising liquidity risk often leads firms to issue additional short-term debt \citep{zhou2023bond}. The latter captures firms' reliance on interest-bearing liabilities, reflecting the burden of interest payments. When both indicators significantly exceed the industry average, the probability of bond default increases markedly.

From a temporal perspective, nearly all clusters exhibit higher attention weights within the Q-1 to Q-2 window, indicating that the model relies heavily on the most recent financial information when forecasting default events. This pattern suggests that firms approaching default often display pronounced financial anomalies, particularly in cash flow, profitability, and balance sheet structure. However, in clusters such as cluster 2 and 3, the debt-to-asset ratio maintains sustained importance even in middle and early periods, implying that the model also captures the process of long-term risk accumulation. This observation is consistent with real-world cases where some firms demonstrate several years of financial deterioration before eventually defaulting.

\begin{figure}[htbp]
    \centering
    \includegraphics[width=0.75\linewidth]{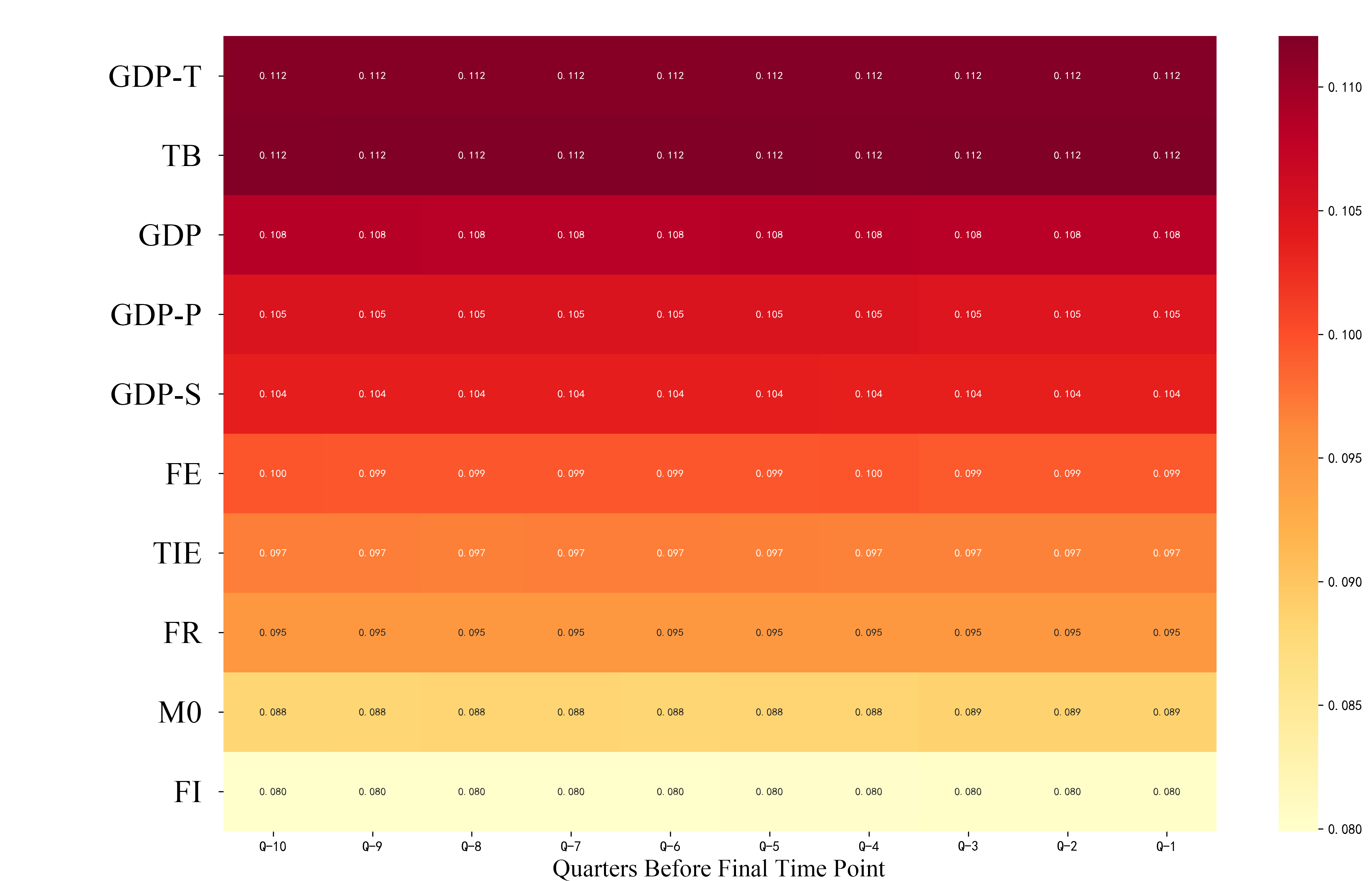}
    \caption{Macroeconomic indicators attention map}
    \label{fig9}
\end{figure}

Figure~\ref{fig9} shows the distribution of attention weights for macroeconomic indicators across time windows. Unlike firm-level indicators with clear temporal variation, macroeconomic indicators remain stable, with nearly uniform weights. This suggests that, due to time lags such as monetary policy effects, the model emphasizes long-term macroeconomic levels rather than short-term fluctuations, consistent with the view that firm-level performance has a more immediate impact on default prediction.

In terms of feature importance rankings, four of the top five macroeconomic indicators are GDP-related variables, namely GDP-tertiary Industry, GDP, GDP-primary industry, and GDP-secondary industry. This indicates that GDP-related measures remain the core reference for the model in capturing the external economic environment, highlighting the strong correlation between overall economic activity, industrial structure shifts, and corporate credit risk. Specifically, GDP are negatively associated with default rates \citep{giesecke2011corporate}. To solve this problem, \citet{onder2023optimal} suggests that GDP-indexed bonds, by linking debt payments to GDP performance, can effectively mitigate default risk.

Of particular note, trade balance and total imports and exports rank second and seventh, respectively, reflecting external shocks such as China-US trade frictions and global supply chain restructuring. These results highlight the trade environment's influence on firms' debt-servicing capacity, especially for export-oriented or import-dependent industries. Greater foreign competition increases firms' reliance on short-term debt \citep{atawnah2023does}, consistent with the preceding discussion that the short-term debt ratio is of high importance. Additionally, official foreign exchange reserves ranks eighth, which aligns with the findings of \citet{cottrell2025us} that US sovereign defaults spread globally through the reserve currency status and lead to interbank bond defaults.

In sum, the feature importance patterns reflected by the attention mechanism are highly consistent with established insights from traditional financial analysis and economic research on the determinants of corporate default.

\section{Conclusion and future work}
\label{sec6}

This study addresses critical gaps in bond default prediction by proposing EMDLOT (Explainable Multimodal Deep Learning for Time-series), a framework that integrates multimodal data, handles irregular time-series, enhances intrinsic interpretability, and enables multi-class classification-filling long-standing voids in both academic research and practical financial risk management.

Empirically, EMDLOT demonstrates superior performance using real-world data from 1,994 Chinese bond-issuing enterprises between 2015 and 2024. Across key metrics critical to financial risk assessment, it outperforms traditional machine learning benchmarks (e.g., Logistic Regression, Random Forest, XGBoost) and deep learning counterparts (e.g., ANN, LSTM). Specifically, its Recall of 0.7547-measuring the ability to detect high-risk "Extended" and "Defaulted" firms-surpasses LSTM's 0.6703 and XGBoost's 0.5806, a critical advantage in practical scenarios where missing defaults (false negatives) carry far greater costs than false positives. Its F1-score (0.7734) and mAP (0.8323) also lead all benchmarks, ensuring balanced precision and reliable probability rankings for multi-class predictions, while maintaining a robust AUC of 0.9435-striking a rare balance between overall discriminative power and sensitivity to minority high-risk classes.

Methodologically, EMDLOT's innovations solve three core challenges in existing models. First, its integration of bond prospectuses as a textual modality-complementing numerical financial and macroeconomic indicators-captures legally binding, pre-issuance risk signals that conventional textual sources (e.g., news, earnings calls) overlook. Ablation tests confirm this: removing textual data reduces Recall by 12.71\% and F1-score by 13.63\%, highlighting prospectuses' unique value in revealing pre-issuance vulnerabilities (e.g., vague risk disclosures linked to higher defaults, as noted in the model's interpretability analysis). Second, the adoption of Time-Aware LSTM (T-LSTM) allows the model to adapt to irregular financial time-series-an issue that cripples standard LSTM and CNN models-by decaying outdated memory based on time intervals, aligning with real-world disclosure practices (e.g., delayed quarterly reports) and reducing reliance on complete data. Third, the combination of soft clustering and multi-level attention mechanisms delivers intrinsic interpretability-a must for financial decision-making-without sacrificing accuracy. Unlike post-hoc tools (e.g., LIME, SHAP) that only supplement "black box" models, EMDLOT's attention mechanisms highlight key signals (e.g., 59.17\% attention to prospectus "Risk Factor" sections, 78.10\% to numerical data) and clustering reveals heterogeneous default drivers (e.g., liquidity shocks vs. long-term leverage), making risk logic transparent to stakeholders.

Practically, EMDLOT bridges the gap between academic models and industry needs. Its three-class classification ("Performing," "Extended," "Defaulted") replaces oversimplified binary frameworks, enabling differentiated risk responses-from monitoring "Extended" firms to restructuring or recovery for "Defaulted" ones. The model's interpretability also supports regulatory compliance and trust-building: attention heatmaps and cluster analyses reveal economically intuitive patterns (e.g., cash flow indicators gain priority in the quarter before default, GDP growth correlates with lower default risk), empowering lenders, investors, and regulators to make informed, actionable decisions.

We expect that  EMDLOT can provide a foundation for expanding multimodal inputs (e.g., dynamic prospectus updates, industry-specific sentiment data) in future research, integrating causal inference to distinguish correlation from causation in risk signals, and adapting the framework to other financial risk tasks (e.g., {delisting prediction, stablecoin risk management, and greenwashing detection}). Ultimately, this work not only delivers a more accurate, interpretable tool for bond default prediction but also sets a new standard for trustworthy, practical financial risk modeling-critical as China's bond market continues to mature and global investors demand greater transparency in credit risk assessment.

\begin{CJK}{UTF8}{gbsn}

\bibliographystyle{refstyle}
\bibliography{references}

\clearpage
\newpage

{\noindent {\Large \textbf{Appendix}}}
\renewcommand{\thesection}{\Alph{section}}
\renewcommand{\thesubsection}{\Alph{section}.\arabic{subsection}}
\renewcommand{\theequation}{\arabic{equation}}
\renewcommand{\thefigure}{\Alph{section}.\arabic{figure}}
\renewcommand{\thetable}{\Alph{section}.\arabic{table}}
\setcounter{section}{0}
\setcounter{equation}{0}
\setcounter{figure}{0}
\setcounter{table}{0}
\numberwithin{equation}{section}
\numberwithin{figure}{section}
\numberwithin{table}{section}

\input{appendix.tex}

\end{CJK}
\end{document}

%% file: appendix.tex
In this online appendix, Appendix A introduces the financial and macroeconomic indicators used in the study, including their abbreviations and summary statistics. Appendix B explains the evaluation metrics—AUC, Recall, F1-score, and mAP—employed to comprehensively assess model performance. Appendix C provides experimental details, such as implementation environment, random seeds, and hyperparameter tuning procedures. Finally, Appendix D presents the soft clustering results, highlighting data heterogeneity and default-related distributional patterns across clusters.

\section{Indicators}
\label{seca}

Table~\ref{taba1} and Table~\ref{taba2} are lists of financial indicators and macroeconomic indicators and their abbreviations, respectively. The abbreviations are mostly used in the figures and tables in this article. Table~\ref{taba3} and Table~\ref{taba4} are their summary statistics.

\begin{table}[htbp]
\centering
\caption{Financial indicators and abbreviations}
\label{taba1}
\begin{tabular}{ll}
\toprule
\multicolumn{1}{c}{Indicators} & \multicolumn{1}{c}{Abbreviation} \\ 
\midrule
Earnings before interest and taxes & EBIT \\
EBIT (estimated) & EBIT-E \\
Earnings before interest, taxes, depreciation and amortization & EBITDA \\
EBITDA (estimated) & EBITDA-E \\
Main operating revenue & MOR \\
Main operating profit & MOP \\
Net profit & NP \\
Main operating profit margin (\%) & MOPM \\
Growth rate of main operating revenue (\%) & MORGR \\
Return on total assets (\%) & ROA \\
Return on equity (\%) & ROE \\
EBITDA to total revenue ratio & EBITDA/TR \\
Operating cash flow to EBITDA & OCF/EBITDA \\
Current ratio & CR \\
Quick ratio & QR \\
Inventory turnover ratio & ITR \\
Debt-to-asset ratio & D/A \\
Short-term debt to total debt & STD/TD \\
Interest-bearing debt to total invested capital & IBD/TIC \\
Cash ratio & CR* \\
Monetary funds to total debt & MF/TD \\
Interest coverage ratio & ICR \\
EBITDA to interest-bearing debt ratio & EBITDA/IBD \\
Net cash flow from operating activities & CFO \\
Net cash flow from investing activities & CFI \\
Net cash flow from financing activities & CFF \\
Net debt & ND \\
Interest-bearing debt & IBD \\
Total debt & TD \\
Net assets & NA \\
Monetary assets & MA \\
Total assets & TA \\
\bottomrule
\end{tabular}
\end{table}

\begin{table}[htbp]
\centering
\caption{Macroeconomic indicators and abbreviations}
\label{taba2}
\begin{tabular}{ll}
\toprule
\multicolumn{1}{c}{Indicators} & \multicolumn{1}{c}{Abbreviation} \\ 
\midrule
Gross domestic product & GDP \\
GDP–primary industry & GDP-P \\
GDP–secondary industry & GDP-S \\
GDP–tertiary industry & GDP-T \\
GDP growth rate (YoY, \%) & GDP YoY \\
GDP–primary industry (YoY, \%) & GDP-P YoY \\
GDP–secondary industry (YoY, \%) & GDP-S YoY \\
GDP–tertiary industry (YoY, \%) & GDP-T YoY \\
GDP growth rate (QoQ, \%) & GDP QoQ \\
Consumer price index & CPI \\
Consumer price index–urban & CPI-U \\
Consumer price index–rural & CPI-R \\
Total imports and exports & TIE \\
Total imports and exports (YoY, \%) & TIE YoY \\
Trade balance & TB \\
Broad money (M2) & M2 \\
Narrow money (M1) & M1 \\
Cash in circulation (M0) & M0 \\
M2 (YoY, \%) & M2 YoY \\
M1 (YoY, \%) & M1 YoY \\
M0 (YoY, \%) & M0 YoY \\
Official foreign exchange reserves & FR \\
End-of-period exchange rate & EoP Rate \\
Total sources of funds of financial institutions & FI \\
Total sources of funds of financial institutions–financial bonds & FI Bonds \\
Business climate index & BCI \\
Industrial climate index & ICI \\
Economist confidence index & ECI \\
Aggregate financing to the real economy (AFRE) & AFRE \\
AFRE (YoY, \%) & AFRE YoY \\
AFRE (QoQ, \%) & AFRE QoQ \\
Consumer confidence index (quarterly) & CCI \\
Fiscal revenue & FR \\
Fiscal revenue (YoY, \%) & FR YoY \\
Fiscal expenditure & FE \\
Fiscal expenditure (YoY, \%) & FE YoY \\
\bottomrule
\end{tabular}
\end{table}

\begin{table}[htbp]
\centering
\caption{Summary statistics of financial indicators}
\label{taba3}
\begin{tabular}{lrrrr}
\hline
Indicators & Min & Max & Mean & S.D. \\
\hline
EBIT        & -5143.23     & 2932.55      & 24.82       & 83.69 \\
EBIT-E      & -11748.06    & 17480.14     & 26.97       & 414.99 \\
EBITDA      & -10129.83    & 41361.11     & 27.00       & 243.25 \\
EBITDA-E    & -11803.18    & 31415.31     & 43.16       & 485.50 \\
MOR         & -44.18       & 38355.71     & 277.55      & 1187.98 \\
MOP         & -4667.48     & 3055.41      & 18.29       & 75.95 \\
NP          & -5241.72     & 1951.07      & 14.29       & 59.68 \\
MOPM        & -4121353.49  & 191048.23    & -111.96     & 22014.18 \\
MORGR       & -1430.73     & 45236383.64  & 3886.31     & 284599.83 \\
ROA         & -128.07      & 141.54       & 3.34        & 4.25 \\
ROE         & -2464.40     & 1345.84      & 4.25        & 26.73 \\
EBITDA/TR   & -714659.50   & 177824.43    & 122.42      & 4138.89 \\
OCF/EBITDA  & -6332.83     & 51669.93     & -0.21       & 237.74 \\
CR          & -142.96      & 411761.36    & 9.32        & 1679.19 \\
QR          & -141.34      & 411761.36    & 8.50        & 1679.18 \\
ITR         & -10528.30    & 100238.71    & 50.63       & 1068.13 \\
D/A         & -11.06       & 1366.79      & 61.35       & 17.00 \\
STD/TD      & -78.06       & 922.16       & 53.92       & 22.20 \\
IBD/TIC     & -2611.12     & 3643.70      & 47.67       & 25.59 \\
CR*         & -55.08       & 376467.25    & 7.04        & 1535.22 \\
MF/TD       & -319.31      & 376467.25    & 6.56        & 1535.22 \\
ICR         & -9601.79     & 88104.91     & 41.57       & 642.22 \\
EBITDA/IBD  & -37609.30    & 153992.58    & 54.98       & 1734.66 \\
CFO         & -4295.16     & 17547.09     & 18.92       & 204.86 \\
CFI         & -13243.79    & 4466.52      & -35.32      & 163.05 \\
CFF         & -2064.83     & 3218.31      & 23.04       & 107.40 \\
ND          & -1897.11     & 144023.98    & 433.11      & 3008.08 \\
IBD         & -96.31       & 156800.34    & 554.78      & 3319.91 \\
TD          & -78.31       & 164378.96    & 938.25      & 3998.32 \\
NA          & -6247.19     & 26747.60     & 426.22      & 1162.38 \\
MA          & -44625.88    & 17247.81     & 134.58      & 494.86 \\
TA          & -69.93       & 175269.09    & 1364.08     & 4729.82 \\
\hline
\end{tabular}
\end{table}

\begin{table}[htbp]
\centering
\caption{Summary statistics of macroeconomic indicators}
\label{taba4}
\begin{tabular}{lrrrr}
\hline
Indicators & Min & Max & Mean & S.D. \\
\hline
GDP          & 147961.80   & 1260582.10  & 616998.90  & 310055.58 \\
GDP-P        & 7770.00     & 89755.20    & 40100.15   & 24698.91 \\
GDP-S        & 58930.60    & 483164.50   & 241051.79  & 121492.74 \\
GDP-T        & 81261.20    & 688238.40   & 335846.96  & 164832.86 \\
GDP YoY      & -6.90       & 18.30       & 5.71       & 3.71 \\
GDP-P YoY    & -3.10       & 8.10        & 3.78       & 1.84 \\
GDP-S YoY    & -9.70       & 24.40       & 5.57       & 4.59 \\
GDP-T YoY    & -5.40       & 15.60       & 6.11       & 3.58 \\
GDP QoQ      & -10.10      & 11.60       & 1.41       & 2.68 \\
CPI          & 100.00      & 104.94      & 101.63     & 1.04 \\
CPI-U        & 99.90       & 104.60      & 101.63     & 0.98 \\
CPI-R        & 99.90       & 105.90      & 101.63     & 1.28 \\
TIE          & 802.14      & 6309.60     & 3099.71    & 1532.37 \\
TIE YoY      & -11.30      & 38.60       & 5.26       & 13.14 \\
TB           & 13.06       & 877.60      & 357.36     & 217.39 \\
M2           & 127533.28   & 309479.82   & 213234.94  & 53334.36 \\
M1           & 33721.05    & 69559.55    & 57345.87   & 9449.94 \\
M0           & 5860.43     & 12183.00    & 8379.74    & 1765.36 \\
M2 YoY       & 6.20        & 13.40       & 9.90       & 1.79 \\
M1 YoY       & -7.40       & 24.70       & 7.11       & 6.93 \\
M0 YoY       & 2.20        & 15.30       & 7.58       & 3.41 \\
FR           & 3009.09     & 3730.04     & 3179.14    & 140.57 \\
EoP Rate     & 6.11        & 7.23        & 6.76       & 0.29 \\
FI           & 1386501.07  & 3676631.17  & 2475762.63 & 657940.30 \\
FI Bonds     & 8836.15     & 152135.12   & 84696.84   & 43872.21 \\
BCI          & 53.24       & 608.11      & 114.85     & 9.32 \\
ICI          & -465.97     & 288.80      & 113.79     & 29.16 \\
ECI          & 60.20       & 151.50      & 92.13      & 21.34 \\
AFRE         & 31058.00    & 145433.00   & 68446.34   & 25922.70 \\
AFRE YoY     & -1366.16    & 372.26      & 11.90      & 35.43 \\
AFRE QoQ     & -63.29      & 1062.07     & 9.97       & 59.27 \\
CCI          & 85.83       & 125.17      & 108.93     & 14.38 \\
FR           & 3640.71     & 21678.40    & 12420.40   & 5112.33 \\
FR YoY       & -14.30      & 24.20       & 4.92       & 8.30 \\
FE           & 3281.53     & 27457.40    & 14581.75   & 6759.99 \\
FE YoY       & -52.32      & 270.79      & 7.13       & 5.88 \\
\hline
\end{tabular}
\end{table}

\clearpage
\section{Evaluation metrics}
\label{secb}

 To comprehensively assess the performance of the multi-class bond default prediction model, this study employs four representative metrics: AUC (Area under the receiver operating characteristic curve), Recall, F1-score, and mAP (Mean average precision). These metrics evaluate the model’s discriminatory power, sensitivity to minority classes, and ranking quality from multiple perspectives, with particular focus on the identification of default categories. Given the severe class imbalance in our dataset, all metrics are calculated using macro averaging rather than weighted averaging, ensuring that each class contributes equally to the overall evaluation.

\subsection{AUC}

AUC quantifies the model’s ability to distinguish between classes across all classification thresholds and is defined as the area under the ROC curve, which plots the True Positive Rate (TPR) against the False Positive Rate (FPR):

\begin{equation}
\mathrm{AUC} = \int_0^1 \mathrm{TPR}(\mathrm{FPR}) \, d\mathrm{FPR}
\end{equation}

where

\begin{equation}
\mathrm{TPR} = \frac{TP}{TP + FN}, \quad \mathrm{FPR} = \frac{FP}{FP + TN}
\end{equation}

To adapt to the multi-class nature of the task, we adopt the One-vs-Rest (OvR) strategy to design AUC. This strategy decomposes the multi-class problem into multiple binary classification problems by treating each class as the positive class against all other classes combined as negative, enabling a more effective measurement of the model’s performance on each individual class. The AUC is then averaged across all classes as:

\begin{equation}
\mathrm{AUC}_{\mathrm{OvR}} = \frac{1}{3} \sum_{i \in \{0,1,2\}} \mathrm{AUC}_{i}
\end{equation}

This approach effectively measures the overall discriminatory capacity of the model, especially suitable for imbalanced multi-class scenarios.

\subsection{Recall, F1-score, and mAP}

To comprehensively assess the performance of the multi-class bond default prediction model, this study focuses on recall, F1-score, and mean average precision (mAP). Recall is of particular importance, as it reflects the model’s ability to correctly identify minority classes such as extended and defaulted bonds. While precision is not directly reported, it is computed alongside recall for the purpose of calculating F1-score and mAP.

For each class \(i\), precision and recall are defined as:

\begin{equation}
\mathrm{Precision}_i = \frac{TP_i}{TP_i + FP_i}, \quad
\mathrm{Recall}_i = \frac{TP_i}{TP_i + FN_i}
\end{equation}

where \(TP_i\), \(FP_i\), and \(FN_i\) denote true positives, false positives, and false negatives for class \(i\), respectively.

The F1-score, as the harmonic mean of precision and recall, is given by:

\begin{equation}
\mathrm{F1-score} = \frac{2 \times \mathrm{Precision}_i \times \mathrm{Recall}_i}{\mathrm{Precision}_i + \mathrm{Recall}_i}
\end{equation}

Macro-averaged metrics across the three classes are calculated as:

\begin{equation}
\mathrm{Precision} = \frac{1}{3} \sum_{i \in \{\mathrm{0,1,2}\}} \mathrm{Precision}_i
\end{equation}

\begin{equation}
\mathrm{Recall} = \frac{1}{3} \sum_{i \in \{\mathrm{0,1,2}\}} \mathrm{Recall}_i
\end{equation}

\begin{equation}
\mathrm{F1-score} = \frac{1}{3} \sum_{i \in \{\mathrm{0,1,2}\}} \mathrm{F1-score_i}
\end{equation}

Mean average precision evaluates the quality of the model’s confidence-ranked predictions. Its calculation logic is based on the precision-recall curve (P-R curve). For each class \(i\), the Average Precision (AP) is computed as the area under the P-R curve:

\begin{equation}
\mathrm{AP}_i = \int_0^1 \mathrm{Precision}(\mathrm{Recall}) \, d\mathrm{Recall}
\end{equation}

The mAP is then obtained by averaging the AP values of all classes:

\begin{equation}
\mathrm{mAP} = \frac{1}{3} \sum_{i \in \{\mathrm{0,1,2}\}} \mathrm{AP}_i
\end{equation}

This metric not only reflects classification accuracy but also captures the ranking quality of predicted probabilities, making it a robust measure for multi-class and imbalanced problems.

AUC, recall, F1-score, and mAP collectively contribute to a more comprehensive evaluation of the model’s performance in complex multi-class tasks. 

\section{Experimental details}
\label{secc}

The model implementation was carried out in Python. All experiments were executed on a NVIDIA GeForce RTX 5070 Ti GPU and an AMD R9-7845HX CPU.

The random seeds recorded in the experiments are as follows: 86, 127, 71, 61, 84, 140, 88, 46, 131, 53.

\begin{figure}[htbp]
    \centering
    \includegraphics[width=1\linewidth]{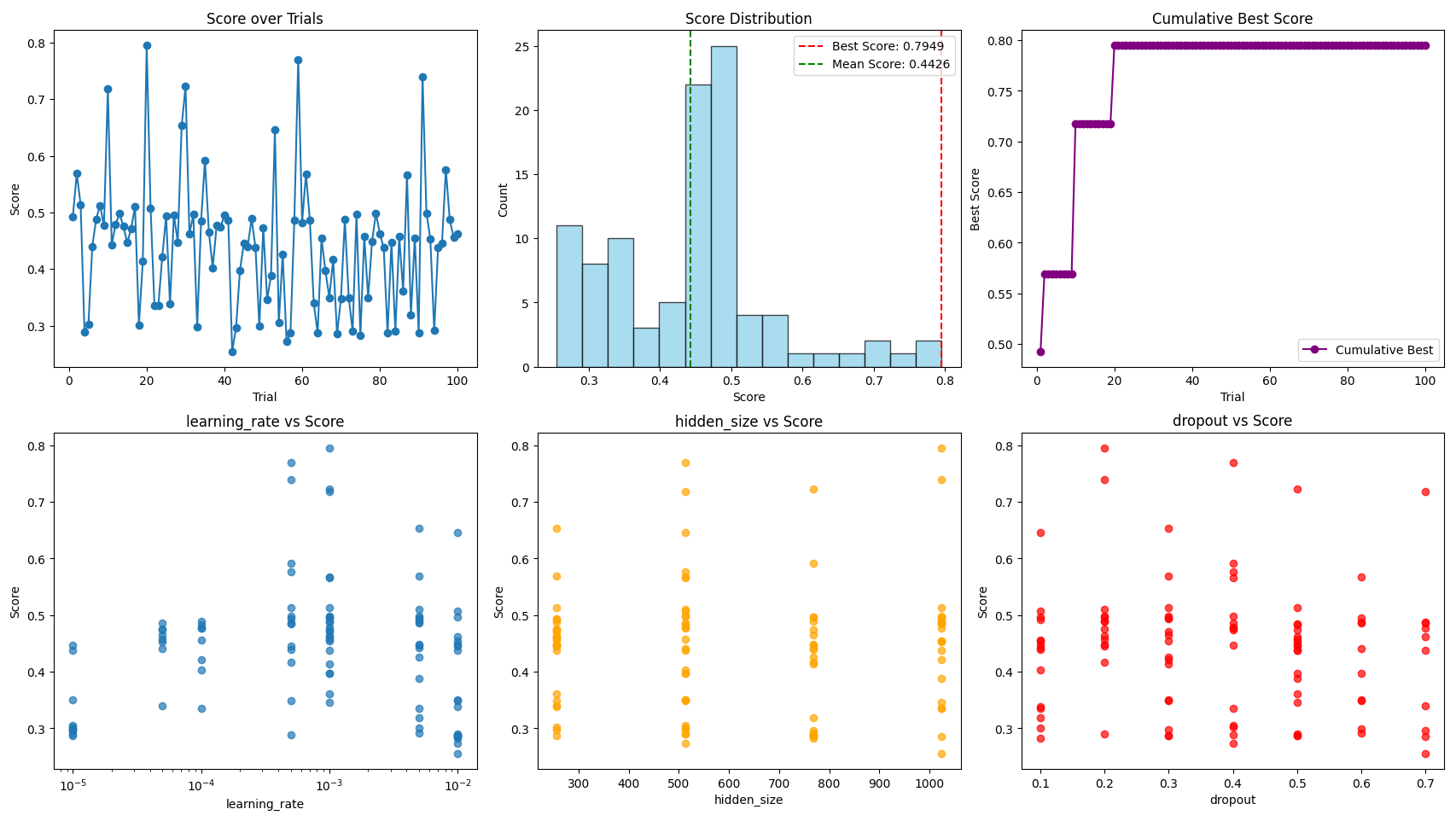}
    \caption{Hyperparameter tuning records with Random Search}
    \label{figc1}
\end{figure}

Figure~\ref{figc1} illustrates a portion of the hyperparameter tuning process with Random Search conducted during the experiments.

\begin{figure}[htbp]
    \centering
    \includegraphics[width=0.75\linewidth]{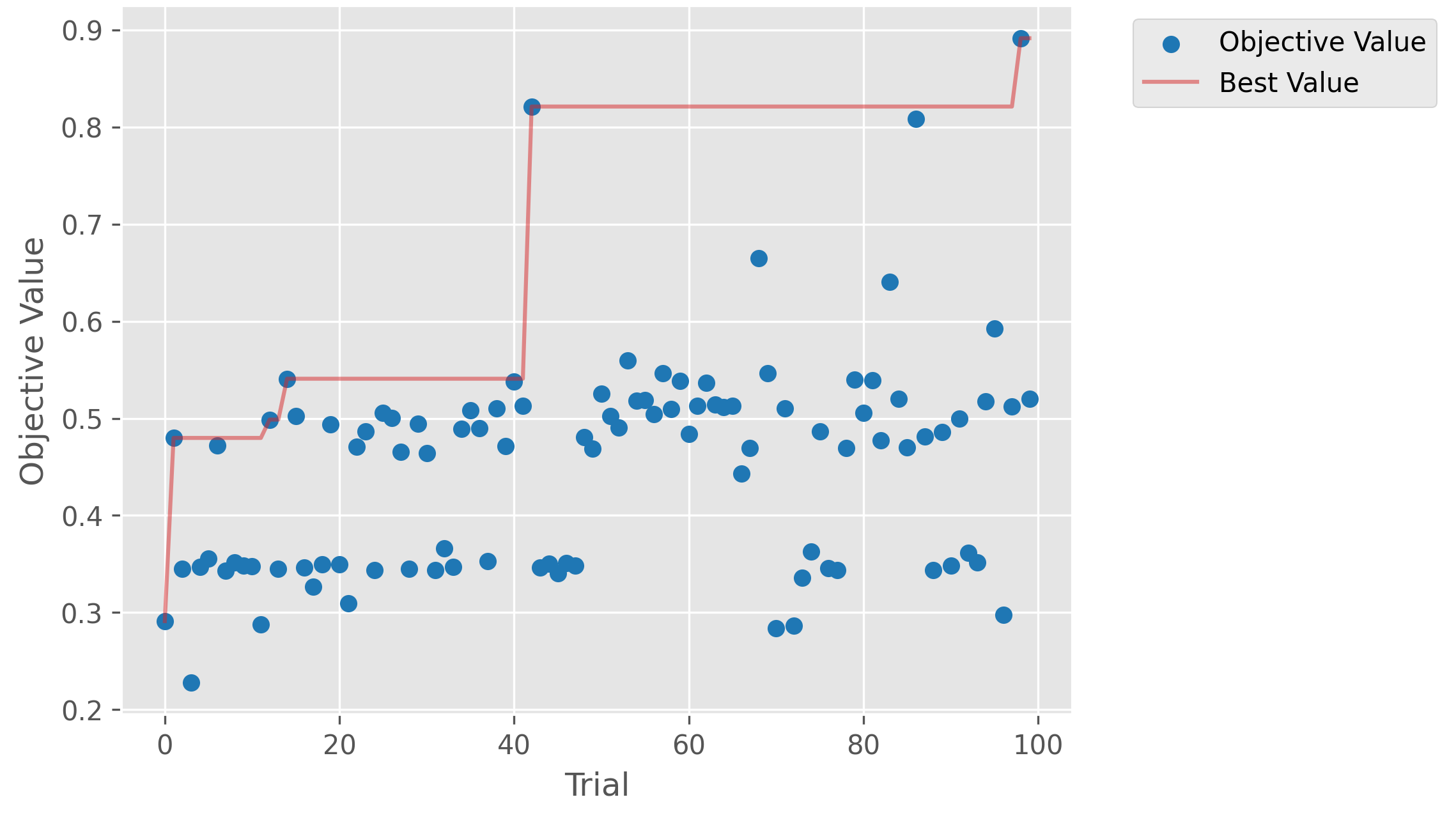}
    \caption{Hyperparameter tuning records with Optuna}
    \label{figc2}
\end{figure}

Figure~\ref{figc2} presents  a typical set of hyperparameter tuning records obtained from Optuna. The Objective Value corresponds to the target value returned by the function objective() in each trial. Since we set direction='maximize', a larger value indicates better performance. In our implementation, the Objective Value is defined as a composite score when the number of predicted classes equals three, and as a penalty score otherwise: 

\begin{equation}
\text{Objective Value} =
\begin{cases}
\text{Composite Score}, & \text{if predicted classes = 3} \\[6pt]
\text{Penalty Score}, & \text{otherwise}
\end{cases}
\end{equation}

The composite score is calculated as a weighted combination of multiple performance metrics:

\begin{equation}
\text{Composite Score} = 0.4 \times \text{Recall} 
+ 0.25 \times \text{F1}
+ 0.25 \times \text{AUROC}
+ 0.1 \times \text{mAP}
\end{equation}

When the number of predicted classes is not three, a penalty is imposed as follows:

\begin{equation}
\text{Penalty Score} = 0.5 \times \text{Composite Score} 
+ 0.3 \times \frac{\text{Unique Preds}}{3.0}
\end{equation}

\begin{figure}[htbp]
    \centering
    \includegraphics[width=0.75\linewidth]{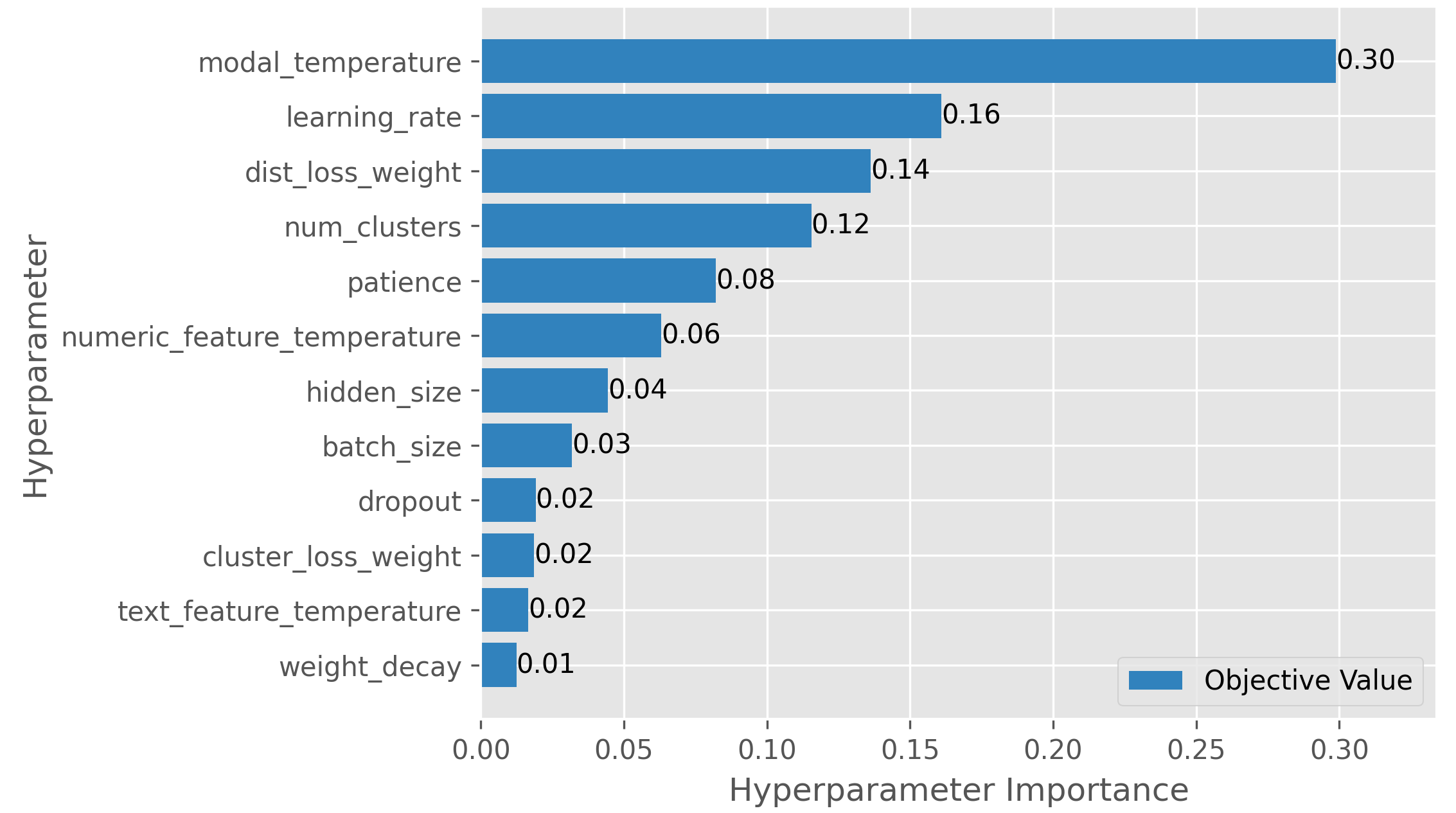}
    \caption{Hyperparameters importance}
    \label{figc3}
\end{figure}

Figure~\ref{figc3} illustrates the relative importance ranking of the hyperparameters.

\clearpage
\section{Soft clustering results}
\label{secd}

In this study, a soft clustering mechanism is introduced as a key intermediate component of the model, aiming to enhance the model’s ability to perceive data heterogeneity while improving the interpretability of results. It allows samples to belong to multiple clusters with varying degrees of membership, thereby capturing the implicit structures and similarity relationships between samples in a more nuanced manner. In practical modeling, this mechanism is applied to reveal the potential distribution patterns of corporate samples in terms of credit risk, facilitating the identification of common characteristics and default tendencies among different types of companies.

Table~\ref{tabd1} presents the soft clustering results from one of the experiments, showing the distribution of sample counts across the three categories (Performing, extended, and defaulted) within each cluster. Overall, there are significant differences in the proportion of samples across the three labels among different clusters. This not only fully reflects the heterogeneity inherent in the data but also validates the effectiveness of the soft clustering mechanism in identifying hierarchical structures within the samples.

For instance, Cluster 7 exhibits a high default risk: it contains 9 extended samples and 3 defaulted samples, both of which are the largest among all clusters. This indicates that enterprises within this cluster are more prone to debt issues and carry strong early warning signals of default. In contrast, Cluster 1 contains no defaulted samples at all, with only 2 extended samples, reflecting a relatively stable credit status. Cluster 2, however, shows a particularly unique pattern: it contains neither extended nor defaulted samples, and its total sample size (93) is significantly smaller than that of other clusters. This suggests that enterprises in this cluster may possess distinct advantages in terms of financial health, information disclosure quality, and other aspects, classifying them as high-quality entities with superior credit ratings.

These striking distributional differences across clusters not only enhance our global understanding of how default-related samples cluster but also lay a theoretical foundation for subsequent models to adopt tailored modeling approaches for distinct subpopulations.

\begin{table}[htbp]
\centering
\caption{Cluster performance}
\label{tabd1}
\begin{tabular}{ccc c}  
\toprule
Cluster & Performing & Extended & Defaulted \\
\midrule
0 & 219 & 2 & 3 \\
1 & 220 & 2 & 0 \\
2 & 93 & 0 & 0 \\
3 & 262 & 3 & 2 \\
4 & 295 & 3 & 2 \\
5 & 288 & 2 & 2 \\
6 & 287 & 3 & 2 \\
7 & 292 & 9 & 3 \\
\bottomrule
\end{tabular}
\end{table}
